\documentclass[preprint,12pt]{elsarticle} 
\usepackage{tikz}
\usepackage{amsmath,bm,algorithm,algorithmic}
\usepackage[mathscr]{euscript}
 \let\mathscr\relax
\usepackage{subfig}
\usepackage{amssymb}
\usepackage{mathrsfs}
\usepackage{graphicx} 
\usepackage{natbib}
\usepackage{lineno,hyperref}
\usepackage{xcolor}
\usepackage{caption}
\usepackage{booktabs}
\usepackage{makecell}
\graphicspath{{}{}}

\begin{document}

\begin{frontmatter}

\title{An efficient optimization based microstructure reconstruction approach with multiple loss functions} 
%
\author[add1]{Anindya Bhaduri}
\author[add1]{Ashwini Gupta}
\author[add2]{Audrey Olivier}
\author[add1]{Lori Graham-Brady\footnote{Corresponding author. Email address: lori@jhu.edu}}
\address[add1]{Department of Civil Engineering, Johns Hopkins University, Baltimore, MD, USA}
\address[add2]{Columbia University, New York, NY, USA}
%
\begin{abstract}
Stochastic microstructure reconstruction involves digital generation of microstructures that match key statistics and characteristics of a (set of) target microstructure(s). This process enables computational analyses on ensembles of microstructures without having to perform exhaustive and  costly experimental characterizations. Statistical functions-based and deep learning-based methods are among the stochastic microstructure reconstruction approaches applicable to a wide range of material systems. In this paper, we integrate statistical descriptors as well as feature maps from a pre-trained deep neural network into an overall loss function for an optimization based reconstruction procedure. This helps us to achieve significant computational efficiency in reconstructing microstructures that retain the critically important physical properties of the target microstructure. 
A numerical example for the microstructure reconstruction of bi-phase random porous ceramic material demonstrates the efficiency of the proposed methodology. We further perform a detailed finite element analysis (FEA) of the reconstructed microstructures to calculate effective elastic modulus, effective thermal conductivity and effective hydraulic conductivity, in order to analyse the algorithm's capacity to capture the variability of these material properties with respect to those of the target microstructure. This method provides an economic, efficient and easy-to-use approach for reconstructing random multiphase materials in 2D which has potential to be extended to 3D structures.
\end{abstract}
\end{frontmatter}
\section{Introduction}
The discovery and development of new materials with targeted properties has always been a central theme for computational material science research, which requires a fundamental understanding of materials behavior and properties across different scales. Stochastic microstructure reconstruction is an important piece of that challenging puzzle. It enables computational generation of microstructures that match the statistically equivalent characteristic(s) of a (set of) target microstructure(s), avoiding the need for exhaustive and costly physical microstructure characterization of every sample to be evaluated.\\
\indent There exists a number of different types of microstructure reconstruction approaches in the current literature. Reconstruction via statistical functions \cite{hazlett1997statistical, rintoul1997reconstruction, graham2008stochastic} is a popular technique applicable to a wide range of materials such as, particulate structures \cite{rintoul1997reconstruction, collins2010three}, chalk \cite{talukdar2002stochastic}, soil \cite{gerke2012description}, sandstone \cite{jiao2008modeling, tang2009pixel}.The reconstruction is usually done via a stochastic optimization method, known as Yeong and Torquato (YT) method \cite{yeong1998reconstructing, torquato1998reconstructing} and tries to ensure that the chosen statistical functions of the target microstructure match closely with that of the reconstructed microstructure. However, if the target image is anisotropic or multiphase or large in size, the convergence of the YT method is especially slow which makes the procedure computationally expensive. 
Physical descriptor-based approaches \cite{breneman2013stalking, xu2014descriptor1, xu2014descriptor2}, on the other hand, involve a relatively cheaper optimization procedure which tries to match the physically meaningful descriptors of the target microstructure to that of the reconstructed one. However, their application is limited to the microstructure reconstruction of crystalline and particulate structures (regular geometries) and not to material systems with irregular geometries. 
Spectral density function (SDF)-based approaches aim to reconstruct statistically equivalent microstructure by generating SDF representations of the target microstructure and matching it to the reconstructed one. The methods \cite{quiblier1984new, levitz1998off, cahn1965phase, yu2017characterization}  are mostly analytical in nature and hence much faster than the optimization-based reconstruction approaches but their application is restricted to isotropic binary materials. Another class of microstructure reconstruction methods involves using deep learning \cite{schmidhuber2015deep, lecun2015deep}, a machine learning approach that can be used amongst other tasks for surrogate modeling \cite{cristianini2000introduction, williams1998prediction, bhaduri2018efficient, bhaduri2018stochastic, bhaduri2020free,bhaduri2020usefulness, bhaduri2020probabilistic} and thus has been implemented successfully for a wide range of classification and regression tasks. Deep learning approaches, in particular convolutional neural networks, are particularly well-suited to handle image data, and have thus received attention lately from the materials research community to process microstructures image data for a variety of tasks \cite{decost2017exploring, lubbers2017inferring}. Deep learning approaches for reconstructions are of two different types: material-system-dependent and material-system-independent approaches. Material-system dependent approaches include work by Cang et al. \cite{cang2017microstructure} which uses a convolutional deep belief network \cite{lee2009convolutional} and work by Li et al. \cite{li2018deep} where a Generative Adversarial Network (GAN) \cite{goodfellow2014generative} model is employed. These approaches train the weights of the network with images specific to a material system and thus need to be retrained for a new material system. Alternatively, transfer learning approaches are material-system-independent and does not require training weights with a set of materials data. Instead, deep learning models, pre-trained using benchmark datasets in the field of computer vision, are used to generate microstructure reconstructions. However, it is noted that, given the fixed pretrained weights, there still may be a need for hyperparameter tuning to get the most optimal results for a given material data set. The work by Li et al. \cite{li2018transfer} is one such approach where a deep convolutional network VGG-19 \cite{simonyan2014very} pretrained on ImageNet \cite{deng2009imagenet} dataset has been used for the reconstruction based on a single given target microstructure. A feature-matching optimization has been performed using a Gram-matrix loss function to generate statistically equivalent microstructures. The reconstruction approach presented in this paper is derived from the work by Li et al \cite{li2018transfer}.\\
\indent In this paper, we propose a modified version of the existing transfer learning approach \cite{li2018transfer}. The modifications include: a) different combinations of Gram matrix layers, b) a weighted combination of the gram matrix (GM) loss components, c) elimination of the potentially costly simulated-annealing based volume fraction matching process, d) inclusion of microstructural descriptor metrics in the overall loss function in the main optimization step. {\color{black}The modifications aim to enhance the efficiency, accuracy as well as interpretability of the proposed modified approach compared to that of the existing approach.  It is noted here that the performance of the proposed approach is assessed via a statistical analysis of both the physics-based descriptors of the microstructure as well as the Finite Element Analysis (FEA) simulated averaged material properties. }The analysis is demonstrated using a true microstructure dataset consisting of $185$ images of size $560$ pixels $\times$ $902$ pixels, corresponding to $185$ successive slices of a porous ceramic microstructure measured by x-ray tomography similar to those found in \cite{nickerson2019permeability, kovci20193d}. The target microstructures are described in section $2$. Section $3$ discusses the denoising of the original target  microstructures and compares the statistical properties between the original and the denoised versions. Section $4$ explains the six microstructural descriptors considered in this study to assess the reconstruction quality. The microstructure reconstruction procedure with different loss function combination and different slices (original and denoised) as the target image is discussed in details in section $5$.
In section $6$, FEA simulated material properties are compared between the original and the reconstructed images. Section $7$ provides a discussion of the additional advantages of the proposed approach. Section $8$ provides the conclusions.
\section{Target microstructures}\label{sec:target microstructures}
\begin{figure}[h!]
\centering
\includegraphics[width=0.7 \textwidth]{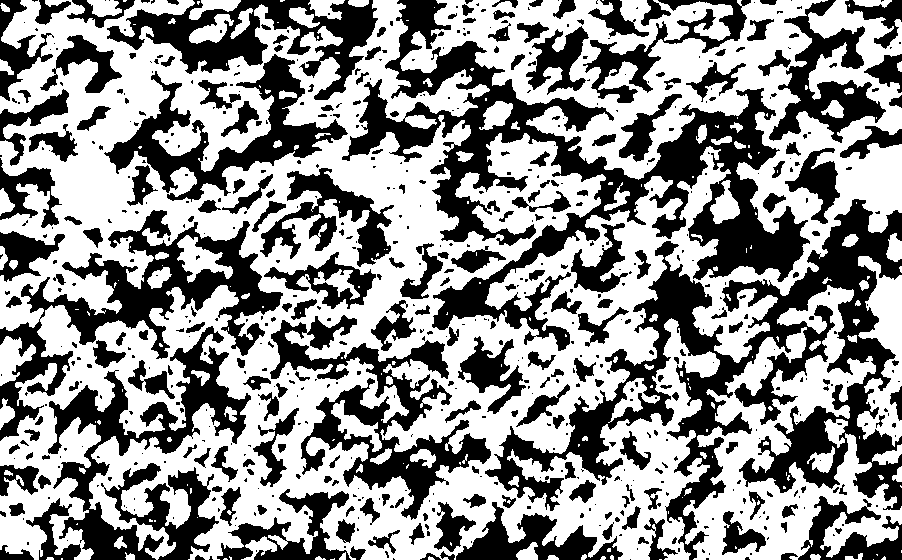}
\caption{Target microstructure (slice $\#$1)}
\label{fig:slice1}
\end{figure}

The target microstructure is that of a two-phase porous ceramic material and there are images of $185$ successive slices of size $560$ pixels $\times$ $902$ pixels. Thus a binary indicator function $I_{(1)}$:$(X,Y) \in \Omega \rightarrow \{0,1\}$ represents the microstructure and is defined as
\[
    I_{(1)}(X,Y) = 
\begin{cases}
    1,              & \text{if } (X,Y) \in \Omega^{(1)}\\
    0,              & \text{if } (X,Y) \in \Omega^{(0)}\
\end{cases}
\]
where $\Omega$ is the material domain, $\Omega^{(0)} \subset \Omega$ is the black solid phase domain and $\Omega^{(1)} \subset \Omega$ is the white porous phase domain such that $\Omega^{(0)} +\Omega^{(1)}  =\Omega$ and $\Omega^{(0)} \cap \Omega^{(1)}=\emptyset$. This is discretized in a computational setting by an array of (0,1) values. The slice $1$ microstructure is shown in figure \ref{fig:slice1}.
\section{Denoising of target microstructure}
\begin{figure}[b!]
  \centering
    \subfloat[slice $\#$1 (original)]{\label{fig:EOI_Sur}\includegraphics[width=0.495\textwidth]{slice1_original.png}}\hfill
      \subfloat[slice $\#$1 (denoised)]{\label{fig:EOI_Sec2}\includegraphics[width=0.495\textwidth]
{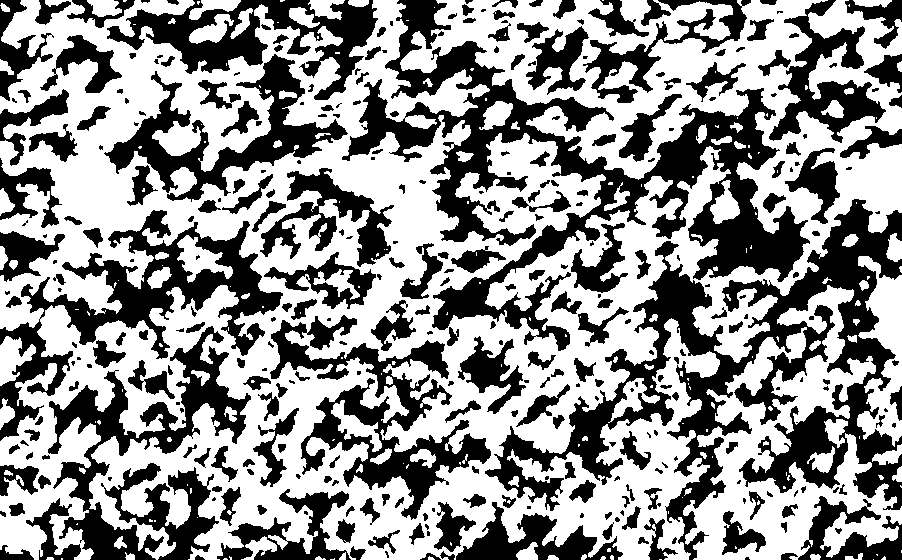}}\hfill
    \subfloat[slice $\#$90 (original)]{\label{fig:EOI_Sec1}\includegraphics[width=0.495\textwidth]
{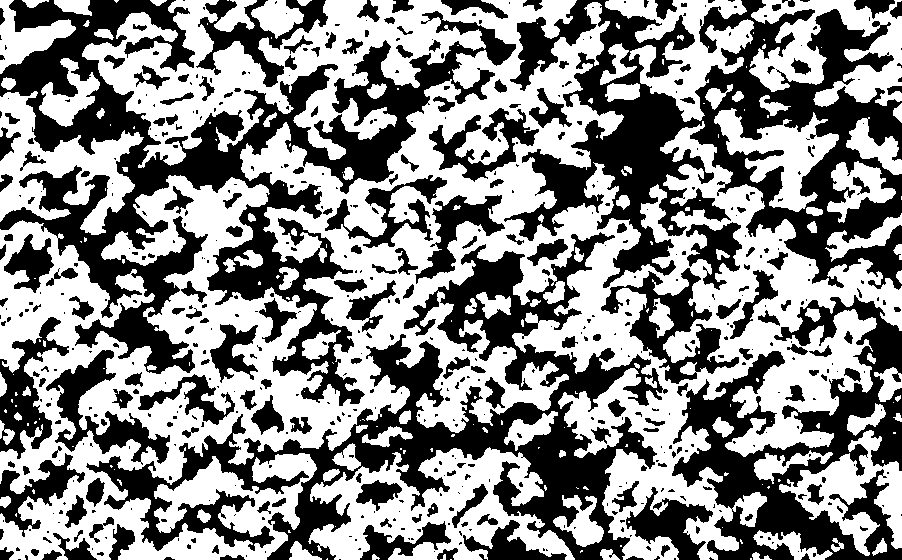}}\hfill
  \subfloat[slice $\#$90 (denoised)]{\label{fig:EOI_Sec2}\includegraphics[width=0.495\textwidth]
{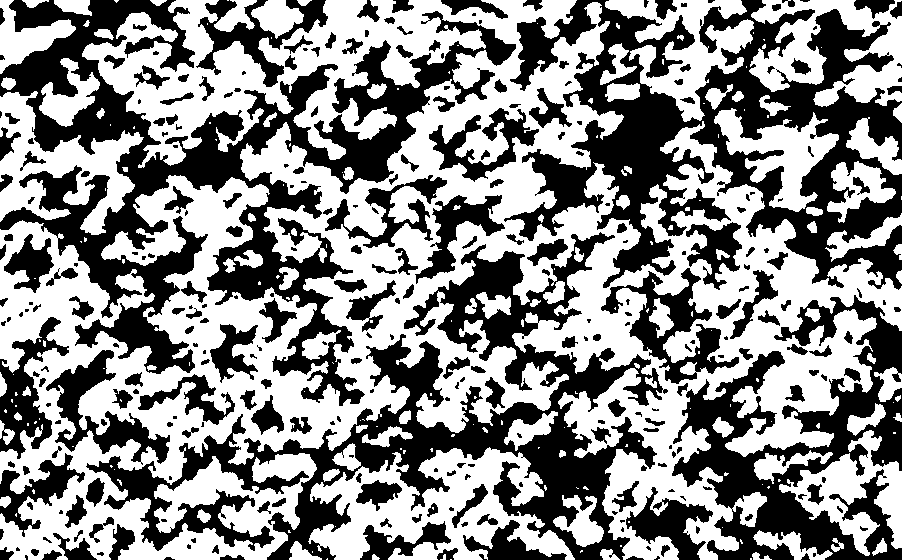}}\hfill
   \subfloat[slice $\#$185 (original)]{\label{fig:EOI_Sec2}\includegraphics[width=0.495\textwidth]
{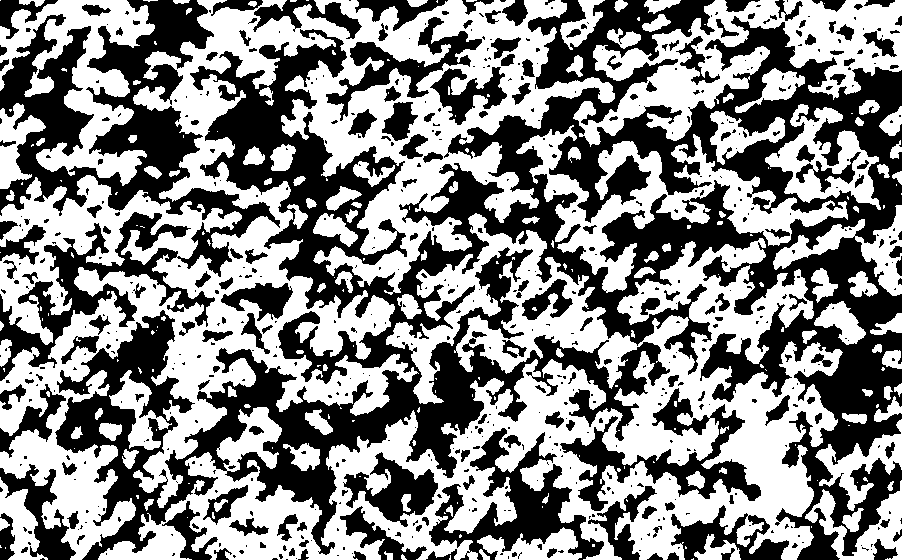}}\hfill
   \subfloat[slice $\#$185 (denoised)]{\label{fig:EOI_Sec2}\includegraphics[width=0.495\textwidth]
{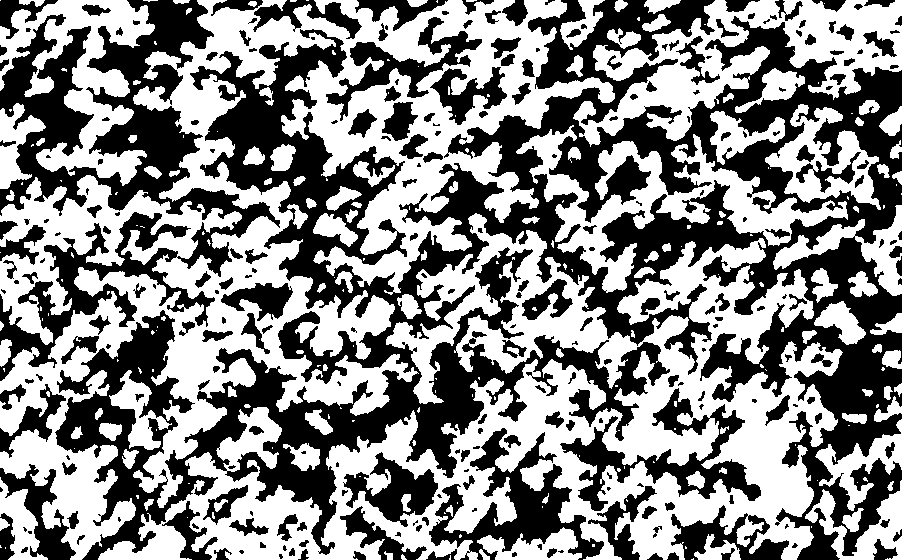}}\hfill
  \caption{Comparison of original and denoised microstructures}
  \label{fig:original_denoised_comparison}
\end{figure}
The need for denoising of the original target microstructure arises from the fact that spurious inclusions of only one or a few pixels appear in the image. These spurious pixels significantly complicate finite element (FE) analyses used to calculate effective properties. While denoising enables coarser discretization of the material domain and more efficient FE analysis, too much denoising can lead to significant changes in the local microstructure. Keeping this in mind, all cluster sizes for the black solid phase less than $10$ pixels are converted to the white porous phase. Figure \ref{fig:original_denoised_comparison} shows a comparison of the original and the corresponding denoised microstructures for slices $\#1$, $\#90$ and $\#185$.
\section{Microstructural descriptors}
\begin{figure}[t!]
\centering
\includegraphics[width=0.75 \textwidth]{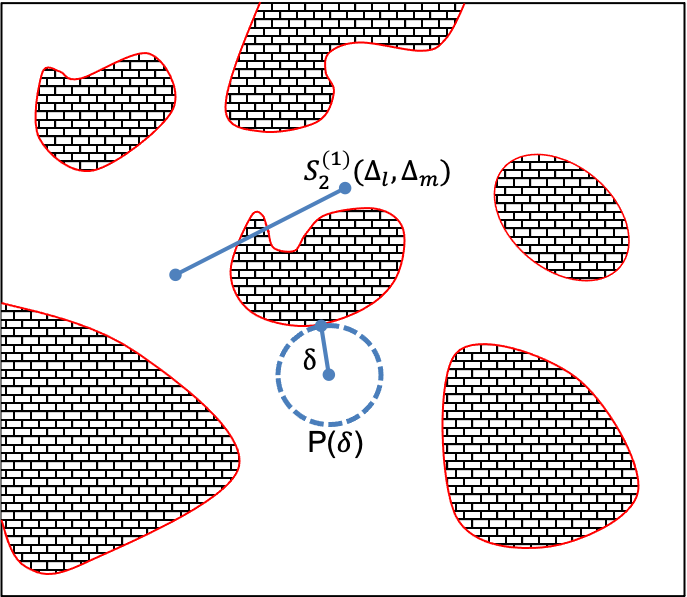}
\caption{A schematic showing different microstructural descriptors in a binary-phase (black solid phase and white porous phase) microstructure.  The two-point correlation function measure for the white phase is represented by a blue solid line.  The red solid lines denote the phase boundaries, the total length of which gives the boundary length.  The pore size distribution function is denoted by $P(\delta)$ where $\delta$ is the pore radius variable.}
\label{fig:micro_descriptors}
\end{figure}

\begin{figure}[t!]
  \centering
    \subfloat[Porosity]{\label{fig:EOI_Sur}\includegraphics[width=0.33\textwidth]{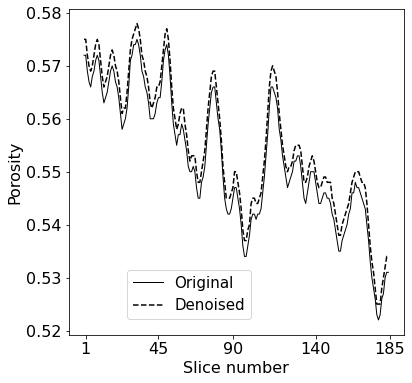}}\hfill
    \subfloat[Correlation length]{\label{fig:EOI_Sec1}\includegraphics[width=0.33\textwidth]
{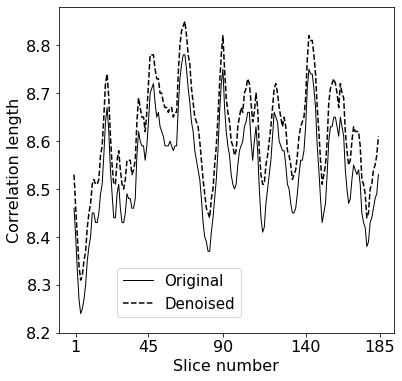}}\hfill
   \subfloat[Specific boundary length]{\label{fig:EOI_Sec2}\includegraphics[width=0.33\textwidth]
{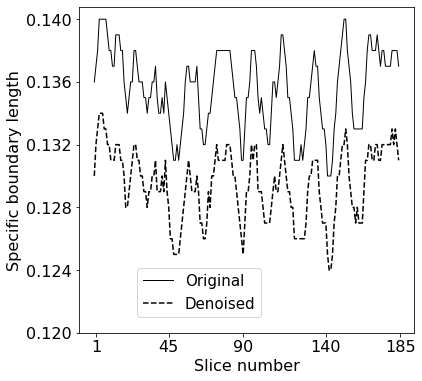}}\hfill
   \subfloat[Mean pore size]{\label{fig:EOI_Sec2}\includegraphics[width=0.33\textwidth]
{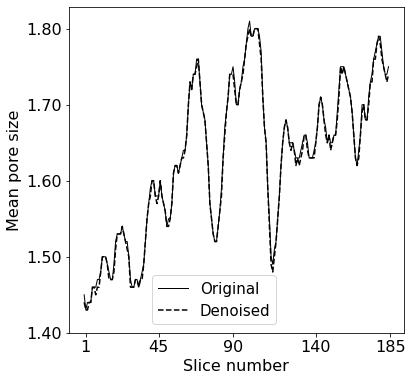}}\hfill
   \subfloat[$90^{th}$-percentile pore size]{\label{fig:EOI_Sec2}\includegraphics[width=0.33\textwidth]
{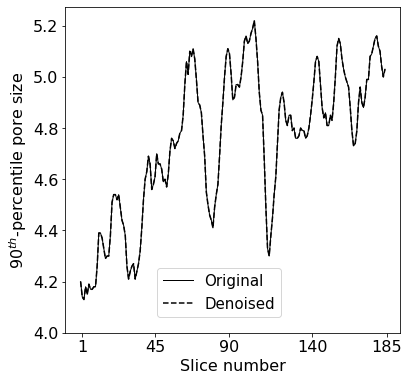}}\hfill
   \subfloat[$97^{th}$-percentile pore size]{\label{fig:EOI_Sec2}\includegraphics[width=0.33\textwidth]
{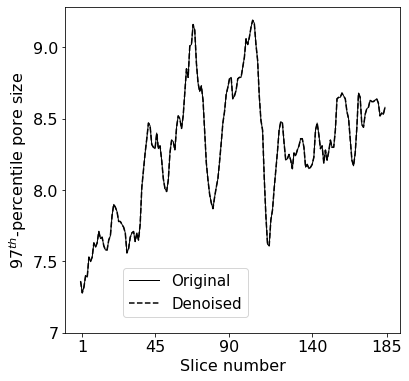}}\hfill
  \caption{Variation of statistical properties across original and denoised slices}
  \label{fig:Comparison_original_denoised_statistics}
\end{figure}

Microstructural descriptors are considered key in determining the effective physical properties of random heterogeneous materials \cite{torquato2002random}. Thus, it is important to match the microstructural descriptors of the reconstructed microstructure with that of the original microstructure. In this study, six such descriptors are considered: porosity, correlation length, specific boundary length, mean pore size, $90^{th}$-percentile pore size and $97^{th}$-percentile pore size. Each of these descriptors are described below and some are shown in figure \ref{fig:micro_descriptors}.\\
\indent Porosity is the fraction of the porous phase in the microstructure given by:
\begin{equation}
\phi = \frac{\sum_{i=1}^{N_x} \sum_{j=1}^{N_y} I_{(1)}(x_i, y_j)}{N_x N_y}
\end{equation}
where $(x_i, y_j)$ is any pixel location in the microstructural domain $\Omega$, while $N_x$ and $N_y$ are the number of pixels along the horizontal and the vertical directions respectively. Figure \ref{fig:Comparison_original_denoised_statistics}(a) shows the porosity of each of the 185 slices of the material. Denoising slightly increases porosity since small regions of the black solid phase are replaced by the white porous phase. \\
\indent While the volume fraction provides information about the microstructural phase at individual points, it does not provide information about the spatial patterns of the microstructure that can be explained to some extent by the two-point probability function. The two-point probability function $S_2^{(1)} (\Delta_l, \Delta_m)$ calculates the probability that any two points $(x_i, y_j)$ and $(x_{i+l}, y_{j+m})$ with a separation vector $(\Delta_l, \Delta_m)$ fall in the white porous phase and is given by:
\begin{align} \label{eq:2pp}
S_2^{(1)} (\Delta_l, \Delta_m) &= P\big{[}I_{(1)}(x_i, y_j)=1, I_{(1)}(x_{i+l}, y_{j+m})=1\big{]} \nonumber \\
&= \frac{\sum_{i=1}^{N_x-l}\sum_{j=1}^{N_y-m}I_{(1)}(x_i, y_j)I_{(1)}(x_{i+l}, y_{j+m})}{(N_x - l)(N_y - m)}
\end{align}
It thus provides a measure of how the two end points of a vector (denoted by a blue solid line in figure \ref{fig:micro_descriptors}) in a phase (in this case, the white porous phase) are correlated.  It is noted that the above description of the two-point probability function similarly applies for the solid phase. In the above definition, it is assumed that the material is spatially stationary (or statistically homogeneous), which may not be valid for materials in which the microstructural statistics vary spatially (e.g., in a functionally graded material).  Correlation length is one scalar measure associated with the two-point probability function, which is calculated by integrating the normalized two-point probability function over the entire domain \cite{vanmarcke1983random} and then taking its square root:
\begin{equation}
\alpha = \sqrt{\frac{S_2^{(1)} (\Delta_l, \Delta_m) - \phi^2}{\phi - \phi^2}}
\end{equation}
Figure \ref{fig:Comparison_original_denoised_statistics}(b) shows the correlation length of each of the 185 slices of the material. It is noted that denoising slightly increases correlation length because it removes small correlation length inclusions.\\
\indent Specific boundary length is the cumulative length of the boundary of separation between the black solid phase and the white porous phase, normalized by the total microstructural area. The cumulative boundary length denoted by red solid lines in figure \ref{fig:micro_descriptors} is obtained by generating the corresponding morphological gradient of the microstructure by taking the difference of its dilation and erosion. The morphological gradient of the microstructure is a matrix of the same size as the microstructural image with non-zero valued pixels along the boundaries of the two phases and zero-valued pixels elsewhere. The cumulative boundary length is thus obtained by counting all the non-zero pixels in the morphological gradient image. Figure \ref{fig:Comparison_original_denoised_statistics}(c) shows the specific boundary length of each of the 185 slices of the material for the original and denoised cases. It is seen that denoising slightly decreases the specific boundary length because the replacement of small clusters of black solid phase by white porous pixels leads to a decrease in the cumulative boundary length.\\
\indent The pore-size probability density function $P(\delta)$ is used to characterize the void or ``pore" space in porous media \cite{prager1963interphase}. The function $P(\delta)$ for isotropic media is defined such that $P(\delta)d\delta$ is the probability that a randomly chosen point in the porous phase domain $\Omega^{(1)}$ lies at a distance between $\delta$ and $\delta + d\delta$ from the nearest point on the pore-solid interface. Thus, $\int_{0}^\infty P(\delta)d\delta = 1$.  The mean pore size is defined as:
\begin{equation}
\langle \delta \rangle = \int_{0}^{\infty} \delta P(\delta)d\delta
\end{equation}
\indent The associated complementary cumulative distribution function $F(\delta)$ is the fraction of pore space that has a pore radius larger than $\delta$ and is given by:
\begin{equation}
F(\delta) = \int_{\delta}^{\infty} P(\delta)d\delta
\end{equation}
$90^{th}$-percentile pore size is defined as the pore size such that $90\%$ of the pore space has a pore radius less than or equal to that pore size, i.e.,$F(\delta)=0.10$. $97^{th}$-percentile pore size, similarly, is defined as the pore size such that $97\%$ of the pore space has a pore radius less than or equal to that pore size, i.e.,$F(\delta)=0.03$.  Figures \ref{fig:Comparison_original_denoised_statistics}(d), \ref{fig:Comparison_original_denoised_statistics}(e) and \ref{fig:Comparison_original_denoised_statistics}(f) show the variation of the mean pore size, the $90^{th}$-percentile pore size and the $97^{th}$-percentile pore size, respectively, across all 185 slices for the original and their corresponding denoised microstructures. It is seen that denoising has very little effect on these three statistical properties of the microstructure.
\section{Microstructure reconstruction}
In this section, the reconstruction procedure and the corresponding results are discussed. The algorithm is an optimization based procedure and different loss function combinations are implemented with an optimal set of hyperparameters for the reconstruction of original slice $\#1$ which is discussed in sections \ref{sec:reconstruction_gm_loss}, \ref{sec:reconstruction_gm_tv_loss} and \ref{sec:reconstruction_gm_tv_2pc_loss}. Section \ref{sec:comparison_three_slices} uses the same set of hyperparameters and compares the reconstruction results of original slice $\#1$ with that of original slices $\#90$ and $\#185$. In section \ref{sec:reconstruction_all_slices}, reconstruction is performed on all the original $185$ slices and their denoised versions with different initializations of the reconstructed image, and the corresponding results are provided. {\color{black}It is noted here that the initial reconstructed image is a white noise color ($3$-channels) image generated by random sampling. Thus the random number generator seed can be changed to generate different initial noisy images that eventually lead to statistically equivalent but distinct reconstructed microstructures. The reconstruction algorithm involves conversion of the $3$-channels representation of the reconstructed color image to the single channel 2-phase representation. This is achieved by a binary $(0/1)$ classification of the 3D color image pixel coordinates based on their euclidean distances from the 3D black pixel coordinate $[0, 0, 0]$ and the 3D white pixel coordinate $[255, 255, 255]$.}
\subsection{Reconstruction with Gram matrix loss}\label{sec:reconstruction_gm_loss}
\begin{figure}[b!]
\centering
\includegraphics[width=0.8 \textwidth]{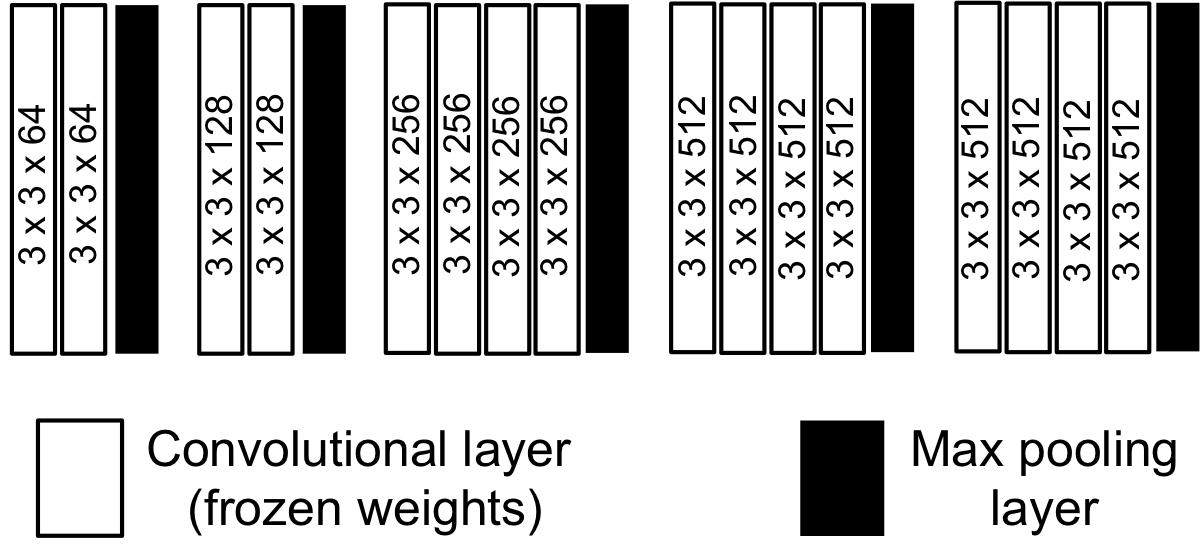}
\caption{VGG-19 network architecture \cite{simonyan2014very}}
\label{fig:VGG19}
\end{figure} 
In this section, a transfer learning based pattern-matching optimization is performed using a pre-trained VGG-19 network shown in figure \ref{fig:VGG19}. {\color{black}VGG-19 is a deep convolutional neural network and in total has $16$ convolutional layers and $5$ max pooling layers, divided into $5$ blocks. The original network, in addition, has $3$ fully connected layers at the end which are excluded here. It is noted that $3$-channels representation is a requirement for the input image to this network \cite{li2018transfer}. {\color{black}Thus, the $2$-phase single channel representation of the original microstructure is converted to a $3$-channels representation by stacking $3$ duplicates of the single channel image.} Any intermediate output from layer operation in a convolutional neural network is called a feature map. Convolutional layers \cite{aghdam2017guide, venkatesan2017convolutional} consists of one or more kernels (matrix of trainable weights) that convolve the input (dot product between the kernel-sized patch of the input and kernel) and pass it on to the next layer. Each convolution layer is followed by a nonlinear activation ReLU (rectified linear unit) \cite{krizhevsky2017imagenet, romanuke2017appropriate} layer which induces non-linearity into the network model. Max pooling \cite{ciregan2012multi, yamaguchi1990neural} uses a max filter for down sampling by selecting the maximum value of the non-intersecting kernel-sized patches of the feature map. Max pooling, in essence, extracts the most prominent features of the previous feature map.} \\
\indent A study has been performed to select the optimal number and combination of VGG-19 network layers for the most accurate reconstruction with respect to the Gram matrix (GM) loss calculation. Gram matrix \cite{gatys2015texture} is a measure of texture in an image and the gram matrix loss here is a measure of the difference between the textures of different intermediate image outputs from the VGG-19 network for the original and the reconstructed microstructure respectively. {\color{black}Following the same notations as in \cite{li2018transfer}, let $F_{pr}^i$ and $\tilde{F}_{pr}^i$ denote the feature maps of the $p^{th}$ filter at position $r$ in layer $i$ for the original and the reconstructed microstructures, respectively. The Gram matrix $G_{pq}^i$ of the original microstructure and $\tilde{G}_{pq}^i$ of the reconstructed microstructure, in layer $i$, is thus defined by the following inner product:
\begin{equation}
G_{pq}^i=\sum_r F_{pr}^i F_{qr}^i
\end{equation}
\begin{equation}
\tilde{G}_{pq}^i=\sum_r \tilde{F}_{pr}^i \tilde{F}_{qr}^i
\end{equation}
The Gram matrix loss for the layer $i$ is given by:
\begin{equation}
E_i = \frac{1}{4N_i^2M_i^2}\sum_{j,k}(G_{jk}^i-\tilde{G}_{jk}^i)^2
\end{equation}
where $N_i$ is the number of filters and $M_i$ is the size of the vectorized
feature maps in layer $i$. The total Gram matrix loss across the selected layers is thus given by:
\begin{equation}
L_G = \sum_i w_i E_i
\end{equation}
where the weight of each layer $i$ is given by $w_i = \frac{min_i \sum_{j,k} (G_{jk}^i)^2}{\sum_{j,k} (G_{jk}^i)^2}$.}
\begin{figure}[t!]
\centering
\includegraphics[width=0.8 \textwidth]{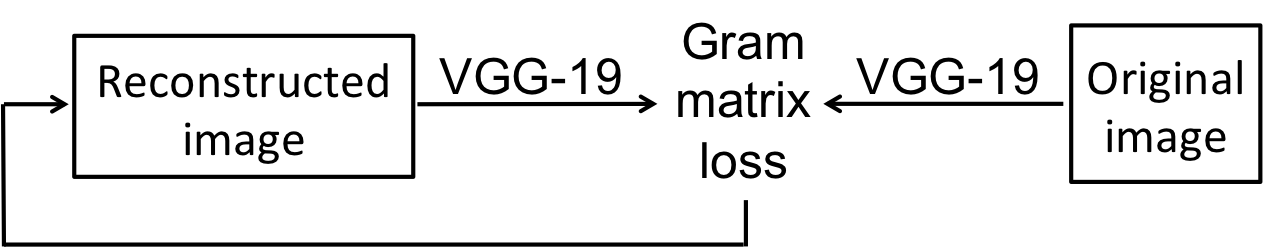}
\caption{Optimization framework with gram matrix loss}
\label{fig:GM_optim}
\end{figure}
The first convolutional layers of each of the 5 blocks of the VGG-19 network as shown in figure \ref{fig:VGG19} is found to be most optimal. The overall optimization framework is shown in figure \ref{fig:GM_optim}. The reconstruction algorithm was run for $50$ different initializations of the reconstructed microstructure. 
\begin{figure}[h!]
  \centering
    \subfloat[Two-point probability function \\ $ \ {} \ {} \ {} \ {} \ {} \ \ \ \ {} $ for the porous phase]{\label{fig:EOI_Sur}\includegraphics[width=0.4\textwidth]{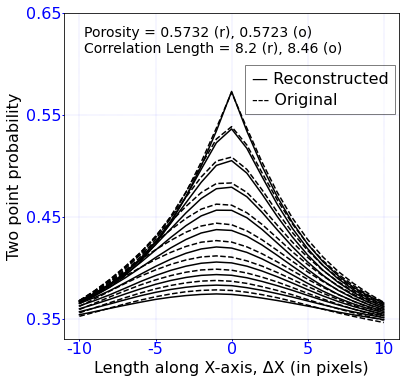}} \hfill
   \subfloat[Pore size complementary cdf]{\label{fig:EOI_Sec2}\includegraphics[width=0.4\textwidth]
{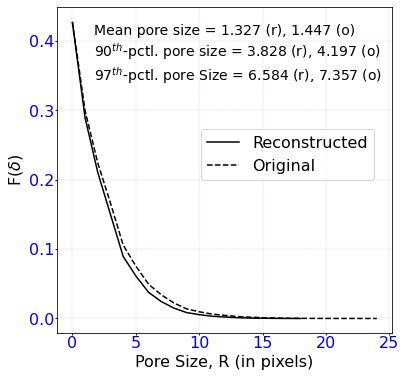}} \hfill
  \caption{Comparison of statistical properties between the original slice $\#1$ and the corresponding reconstructed microstructure using the gram matrix loss function}
  \label{fig:properties_slice1_gmloss}
\end{figure}
Figure \ref{fig:properties_slice1_gmloss} compares the two-point probability function and the pore size cumulative density function (CDF) for the original slice $\#1$ microstructure and the reconstructed image corresponding to one of the $50$ initializations. 
\begin{figure}[t!]
\centering
\includegraphics[width=0.75 \textwidth]{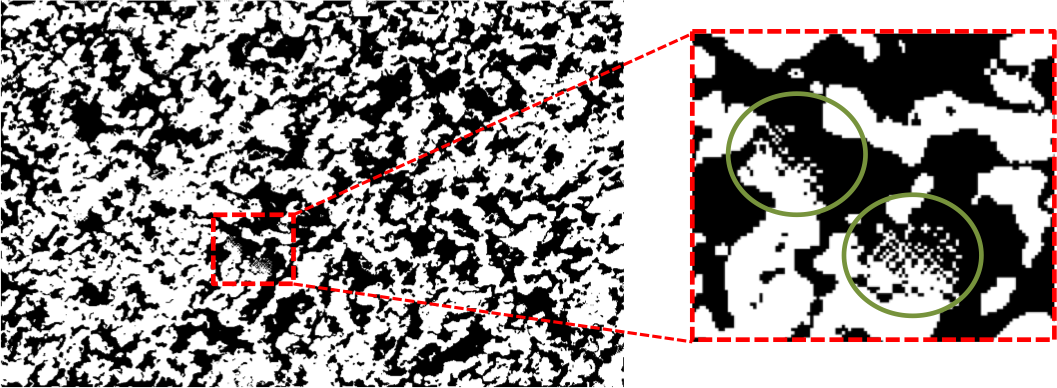}
\caption{Reconstructed microstructure (checkerboard patterns) using gram matrix loss function based optimization}
\label{fig:GM_image}
\end{figure}
Figure \ref{fig:GM_image} shows the reconstructed image corresponding to the original slice $\#1$ image. Although the statistical property estimates of the reconstructed image are close to that of the target slice $\#1$, on visual inspection it is found that spurious checkerboard patterns are present in the reconstructions as seen in figure \ref{fig:GM_image}.
\subsection{Reconstruction with GM and TV loss}\label{sec:reconstruction_gm_tv_loss}
\begin{figure}[b!]
\centering
\includegraphics[width=0.8 \textwidth]{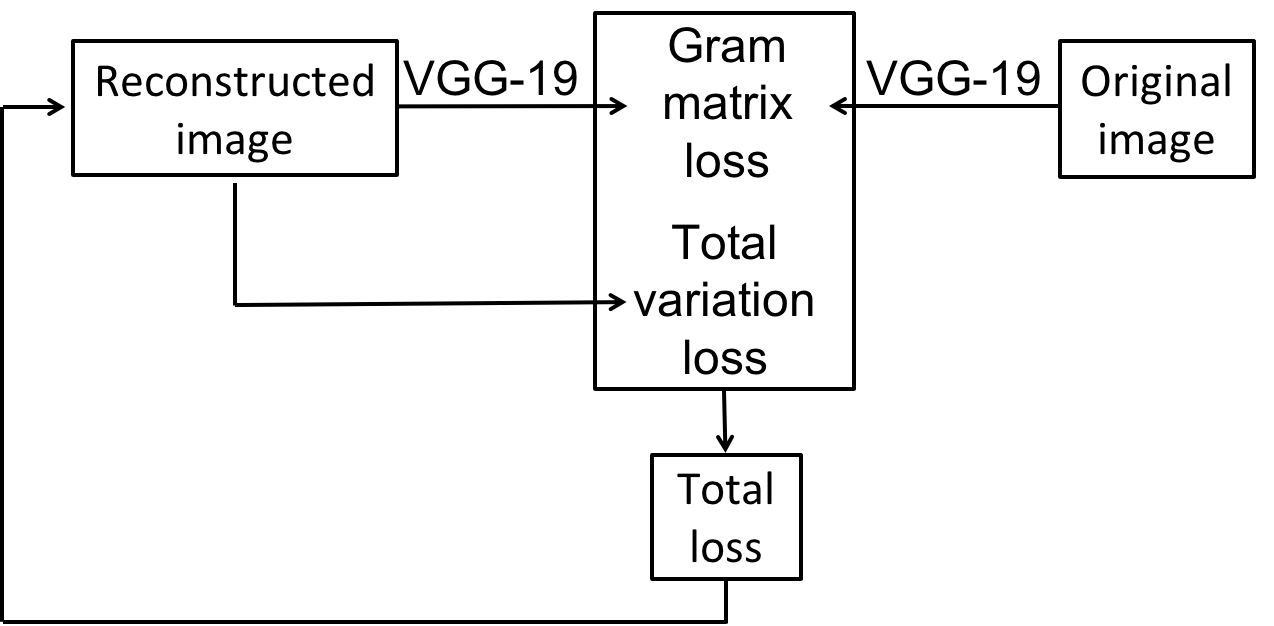}
\caption{Optimization framework with a combination of gram matrix loss and total variation loss}
\label{fig:GM_TV_optim}
\end{figure}
\begin{figure}[t]
\centering
\includegraphics[width=0.5 \textwidth]{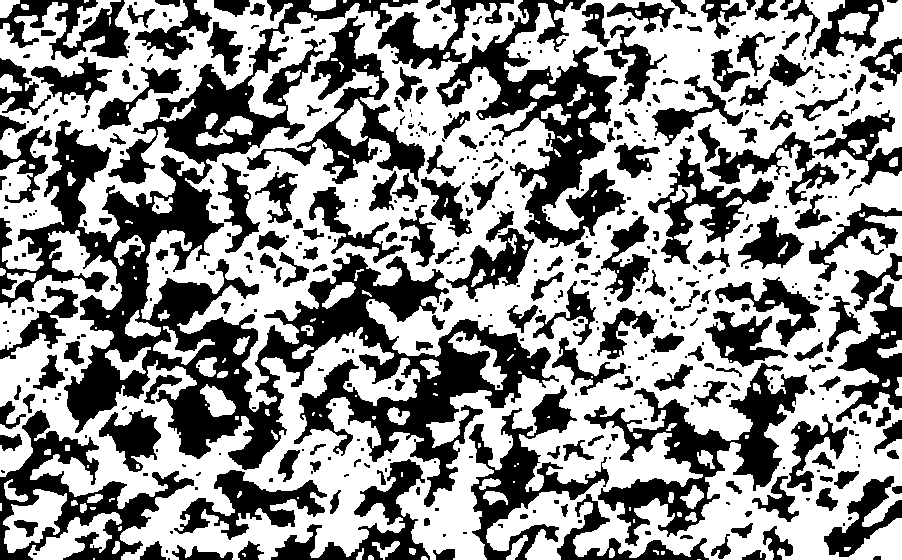}
\caption{Reconstructed microstructure (no checkerboard patterns) using a combination of gram matrix loss and total variation loss for optimization}
\label{fig:GM_TV_image}
\end{figure}
To address the checkerboard patterns, total variation (TV) loss between the target image and the reconstructed image is included in the optimization framework as shown in figure \ref{fig:GM_TV_optim}. The total variation for an image is defined as the sum of the absolute differences in the pixel values between neighboring pixels in that image. It is a measure of noise in an image that is sensitive to features like checkerboard patterns. {\color{black}The total variation loss is the anisotropic version of the total variation norm \cite{rudin1992nonlinear} and is given by:
\begin{equation}
L_V = \sum_k \sum_{i,j} |y_{i+1,j,k}-y_{i,j,k}| +  |y_{i,j+1,k}-y_{i,j,k}|
\end{equation}
where $y_{i,j,k}$ is the ${(i,j,k)}$-th pixel value of the 3D representation of the reconstructed image.}
Figure \ref{fig:GM_TV_image} shows the reconstructed image corresponding to one of the initializations, and it is evident that the inclusion of the total variation loss eliminates the checkerboard patterns found in the reconstructed images in section \ref{sec:reconstruction_gm_loss}. It is also noted that similar observations are found in all the other initialization cases. {\color{black}The total loss $L_T$ in this case is given by:
\begin{equation}
L_T = w_G L_G +w_V L_V
\end{equation}
where $w_G = 1$ and $w_V = 1$e$-6$ are the weight combinations that seem to produce the best results using `trial and error' approach.}
Table \ref{tab:GM_GMTV_mean_comp} shows the mean error in the statistical properties of interest calculated over $50$ different initialization cases for the reconstruction with the combined ``GM+TV" loss and is compared to the corresponding results with only the GM loss discussed in section \ref{sec:reconstruction_gm_loss}. It is seen that the mean error for almost all the statistical properties improves for the ``GM+TV" loss case compared to the GM loss case. Although the error in the porosity is very low in both cases, the porosity matches slightly better on an average for the GM loss case.

%
%
\begin{table}[h!]
  \centering
  \caption{Error in reconstruction with GM loss and ``GM+TV" loss}
  \label{tab:GM_GMTV_mean_comp}
  \begin{tabular}{c|c|c}
    \toprule
    \multicolumn{1}{c|}{} & \makecell{GM loss} & \makecell{GM+TV loss}\\
   \midrule
    \makecell{Mean error in porosity} & 0.0027 & 0.0048  \\
      \makecell{Mean error in correlation length} & 0.2923 & 0.0580 \\
      \makecell{Mean error in specific boundary length} & 0.0144 & 0.0050\\
      \makecell{Mean error in mean pore size} & 0.1225 & 0.0228\\
      \makecell{Mean error in 90-th percentile pore size} & 0.3846 & 0.0947\\
      \makecell{Mean error in 97-th percentile pore size} & 0.8641 & 0.1953 \\
    \bottomrule
  \end{tabular}
\end{table}
\subsection{Reconstruction with GM and TV and 2PP loss}\label{sec:reconstruction_gm_tv_2pc_loss}
\begin{figure}[b!]
\centering
\includegraphics[width=0.75 \textwidth]{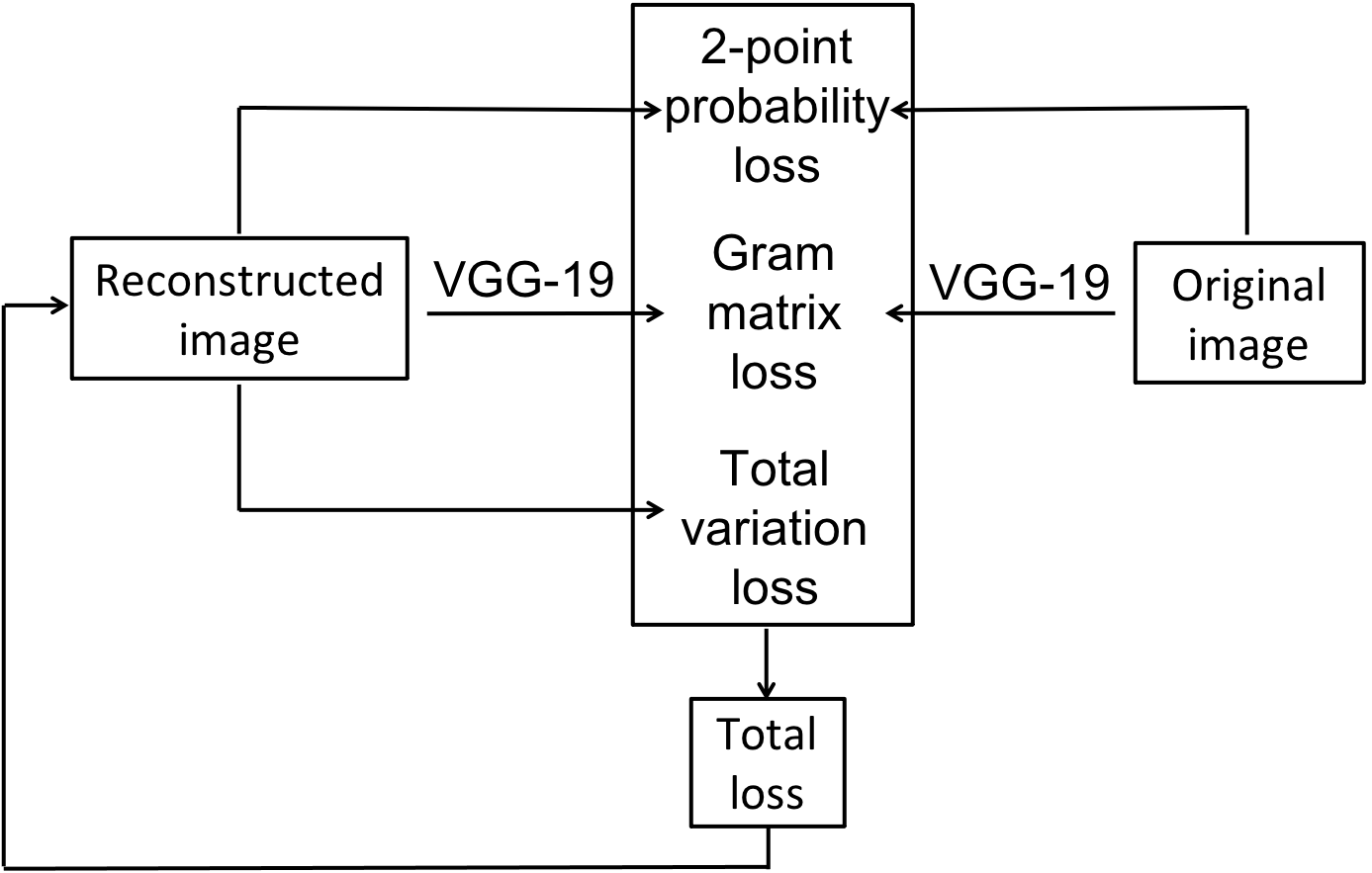}
\caption{Optimization framework with a combination of gram matrix loss, total variation loss and two-point correlation function loss}
\label{fig:GM_TV_2PC_optim}
\end{figure}
\begin{figure}[t!]
  \centering
    \subfloat[Porosity]{\label{fig:EOI_Sur}\includegraphics[width=0.33\textwidth]{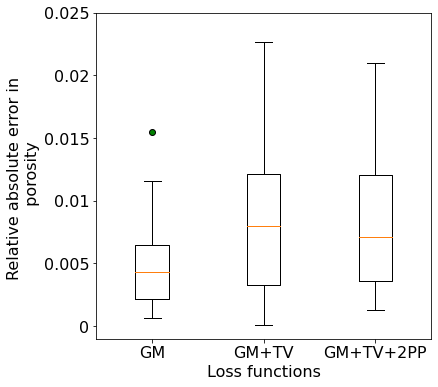}}\hfill
    \subfloat[Correlation length]{\label{fig:EOI_Sec1}\includegraphics[width=0.325\textwidth]
{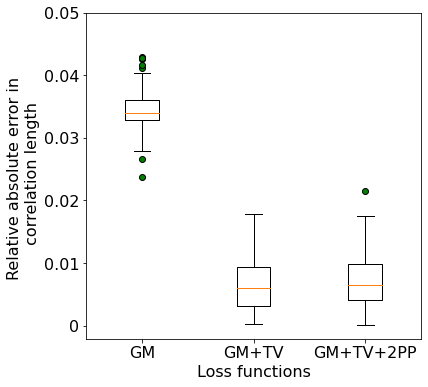}}\hfill
   \subfloat[Specific boundary length]{\label{fig:EOI_Sec2}\includegraphics[width=0.33\textwidth]
{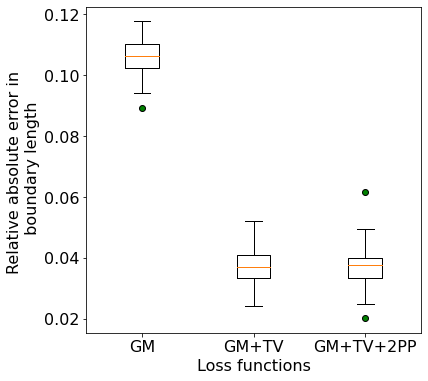}}\hfill
   \subfloat[Mean pore size]{\label{fig:EOI_Sec2}\includegraphics[width=0.33\textwidth]
{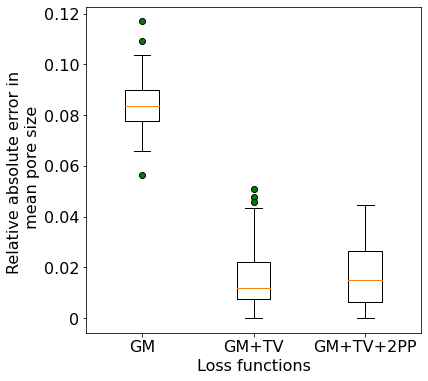}}\hfill
   \subfloat[$90^{th}$-percentile pore size]{\label{fig:EOI_Sec2}\includegraphics[width=0.33\textwidth]
{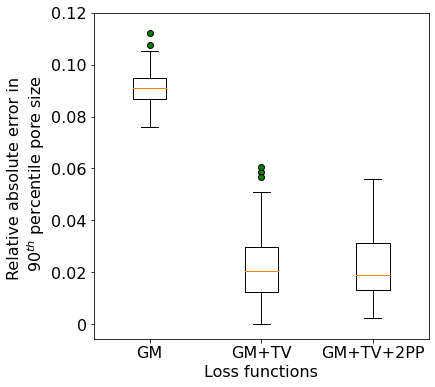}}\hfill
   \subfloat[$97^{th}$-percentile pore size]{\label{fig:EOI_Sec2}\includegraphics[width=0.33\textwidth]
{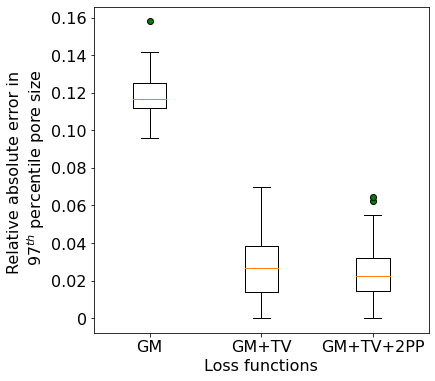}}\hfill
  \caption{Comparison of error in statistical functions for the three different loss function cases}
  \label{fig:comp_props_slice1_allLosses}
\end{figure}

In this section, we try to improve the reconstruction quality further by including another loss function in the optimization, the two-point probability (2PP) loss $L_P$. {\color{black}This loss is defined as the mean squared error between the 2-point probability functions [refer to eq$.$ (\ref{eq:2pp})] of the original and the reconstructed microstructures. }The two-point probability is sensitive to the spatial distribution of a given phase. The overall optimization framework is shown in figure \ref{fig:GM_TV_2PC_optim}. {\color{black}The total loss $L_T$ in this case is given by:
\begin{equation}
L_T = w_G L_G +w_V L_V +w_PL_P
\end{equation}
where $w_G = 1$, $w_V = 1$e$-6$ and $w_P=1$e$8$ are the weight combinations that seem to produce the best results using `trial and error' approach.} It is noted that a loss function corresponding to any other physical property of interest can be added to the optimization framework in a similar fashion. Figure \ref{fig:comp_props_slice1_allLosses} shows the box plot of the errors in the statistical properties of interest corresponding to each of the loss function cases: ``GM", ``GM+TV" and ``GM+TV+2PP" for the reconstruction of original slice $\#1$. The errors are calculated over $50$ different initializations of the reconstructed image. 
It is clear from the box plots that, with the exception of the very small porosity errors, all the other statistical properties match more accurately for the ``GM+TV" and ``GM+TV+2PP" loss cases compared to the ``GM" loss case. However, the performance of the ``GM+TV" and ``GM+TV+2PP" loss cases are very similar to each other and the addition of the two-point probability loss function to the optimization framework does not improve the reconstruction results significantly, which is evident from the mean and coefficient of variation (CV) values of the absolute error in the properties shown in table \ref{tab:GMTV_GMTV2PP_mean_comp}.
\begin{table}[h!]
  \centering
  \caption{Error in reconstruction with ``GM+TV" loss and ``GM+TV+2PP" loss}
  \label{tab:GMTV_GMTV2PP_mean_comp}
  \begin{tabular}{c|c|c|c|c}
    \toprule
    \multicolumn{1}{c|}{} & \multicolumn{2}{c||}{\makecell{GM+TV loss}}  & \multicolumn{2}{c}{\makecell{GM+TV+2PP loss}}\\
    \cmidrule{2-5}
    \multicolumn{1}{c|}{} & Mean & \multicolumn{1}{c||}{CV} & Mean & CV \\
    \midrule
    \makecell{Error in porosity} & 0.0048  & \multicolumn{1}{c||}{0.6402} & 0.0046 & 0.6169\\
    \makecell{Error in correlation length} & 0.0580 & \multicolumn{1}{c||}{0.6698} & 0.0615 & 0.6437\\
    \makecell{Error in specific boundary length} & 0.0050 & \multicolumn{1}{c||}{0.1587} & 0.0050& 0.1867\\
      \makecell{Error in mean pore size} & 0.0228 & \multicolumn{1}{c||}{0.7957} & 0.0247 &0.7321\\
      \makecell{Error in 90-th percentile pore size} & 0.0947 & \multicolumn{1}{c||}{0.6825} & 0.0992 & 0.6295\\
      \makecell{Error in 97-th percentile pore size} & 0.1953 & \multicolumn{1}{c||}{0.6432} & 0.1873 & 0.6288\\
    \bottomrule
  \end{tabular}
\end{table}
\subsection{Performance comparison with other slices} \label{sec:comparison_three_slices}
\begin{figure}[b!]
  \centering
    \subfloat[Porosity]{\label{fig:EOI_Sur}\includegraphics[width=0.33\textwidth]{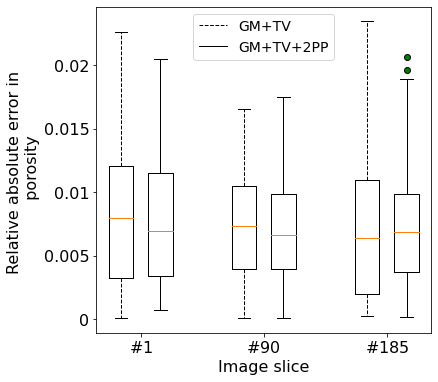}}\hfill
    \subfloat[Correlation length]{\label{fig:EOI_Sec1}\includegraphics[width=0.33\textwidth]
{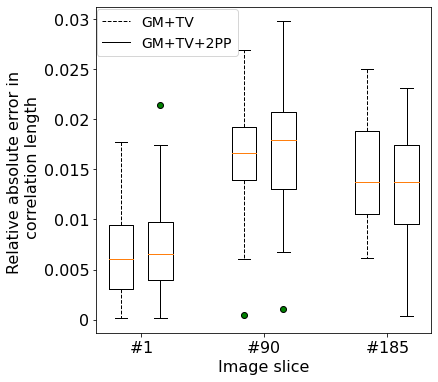}}\hfill
   \subfloat[Specific boundary length]{\label{fig:EOI_Sec2}\includegraphics[width=0.33\textwidth]
{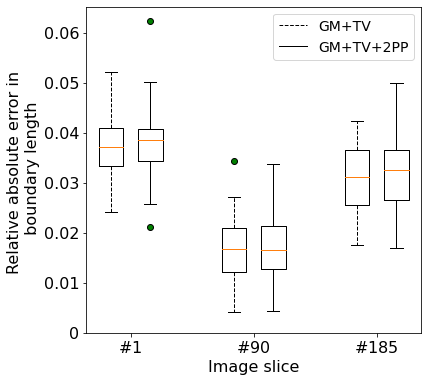}}\hfill
   \subfloat[Mean pore size]{\label{fig:EOI_Sec2}\includegraphics[width=0.33\textwidth]
{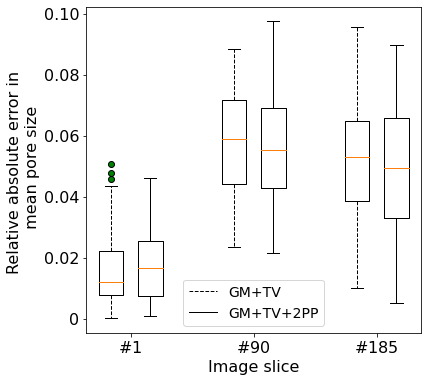}}\hfill
   \subfloat[$90^{th}$-percentile pore size]{\label{fig:EOI_Sec2}\includegraphics[width=0.33\textwidth]
{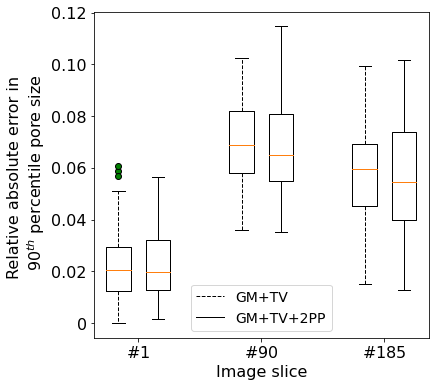}}\hfill
   \subfloat[$97^{th}$-percentile pore size]{\label{fig:EOI_Sec2}\includegraphics[width=0.33\textwidth]
{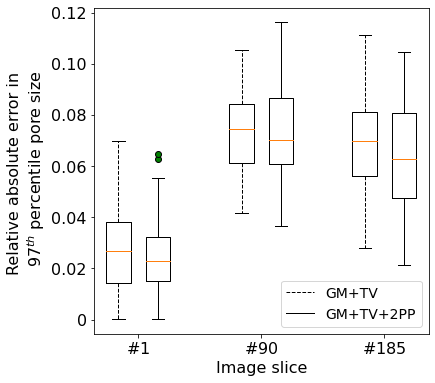}}\hfill
  \caption{Errors in the statistical properties across three different slices (1, 90, 185) for GM+TV and GM+TV+2PP loss}
  \label{fig:comp_3slices_gmtv_gmtv2pp}
\end{figure}
In the previous sections, reconstructions were performed with original slice $\#1$ as the target microstructure. In this section, with the same hyperparameter settings for the ``GM+TV" and ``GM+TV+2PP" loss cases as for slice $\#1$-based reconstruction, the reconstruction algorithms are run with original slices $\#90$ and $\#185$ as the target microstructure over $50$ initializations. These runs are performed to verify whether the optimal hyperparameter settings for the slice $\#1$-based reconstruction works well for the reconstruction on other slices too. Slices $\#90$ and $\#185$ are chosen as representative slices for that study. 
Figure \ref{fig:comp_3slices_gmtv_gmtv2pp} shows the performance comparison in the statistical properties for the ``GM+TV" and ``GM+TV+2PP" loss cases when the target microstructures are slices  $\#1$, $\#90$ and $\#185$ respectively. It is seen that the relative absolute errors in the statistical properties are in similar orders of magnitude across the three different slices for either of the loss cases. For each of the three target slice-based reconstructions, the reconstruction quality for both the loss cases are very similar to each other.
\subsection{Reconstruction on all slices}\label{sec:reconstruction_all_slices}
\begin{figure}[b!]
  \centering
  \subfloat[Porosity]{\label{fig:EOI_Sur}\includegraphics[width=0.33\textwidth]{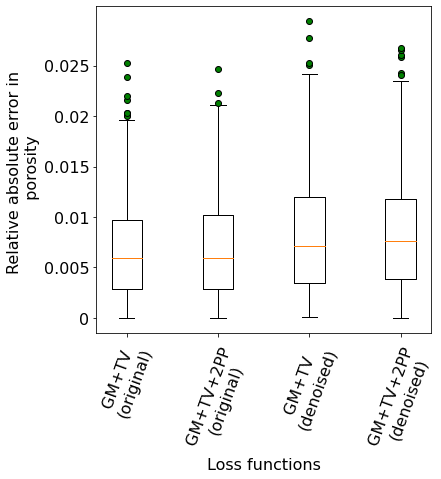}}\hfill
  \subfloat[Correlation length]{\label{fig:EOI_Sec1}\includegraphics[width=0.33\textwidth]
{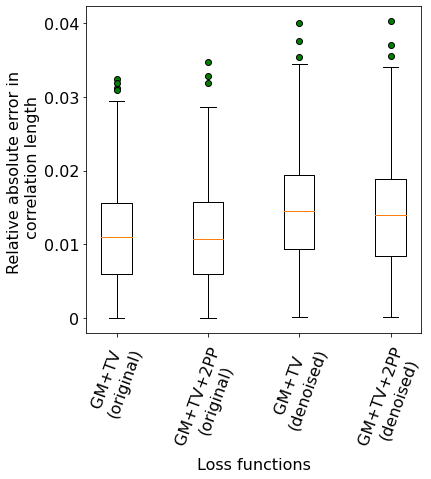}}\hfill
 \subfloat[Specific boundary length]{\label{fig:EOI_Sec2}\includegraphics[width=0.33\textwidth]
{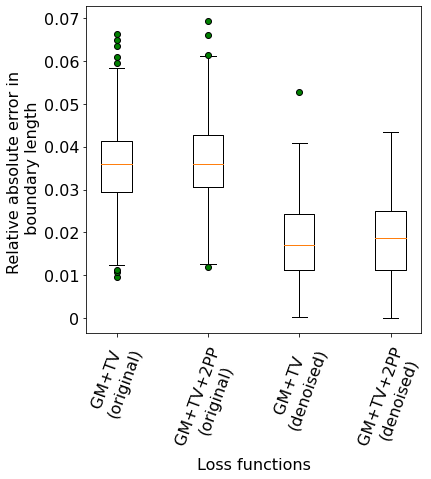}}\hfill
 \subfloat[Mean pore size]{\label{fig:EOI_Sec2}\includegraphics[width=0.33\textwidth]
{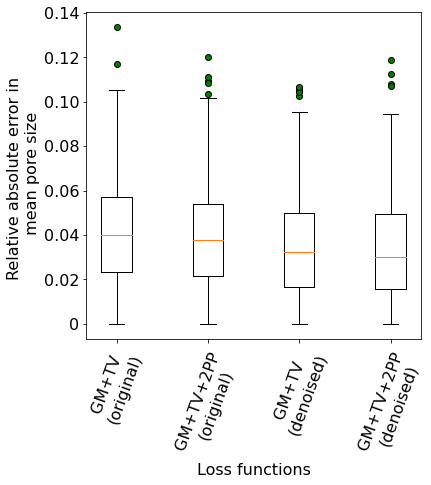}}\hfill
 \subfloat[$90^{th}$-percentile pore size]{\label{fig:EOI_Sec2}\includegraphics[width=0.33\textwidth]
{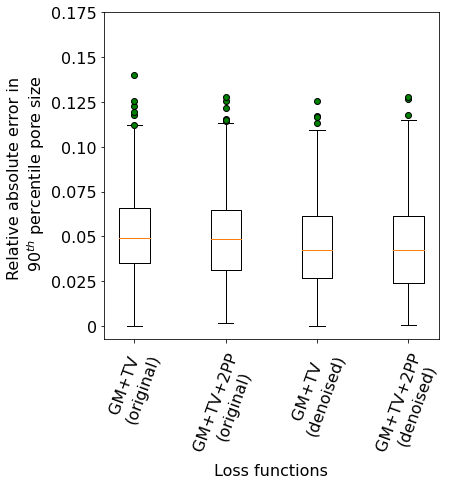}}\hfill
\subfloat[$97^{th}$-percentile pore size]{\label{fig:EOI_Sec2}\includegraphics[width=0.33\textwidth]
{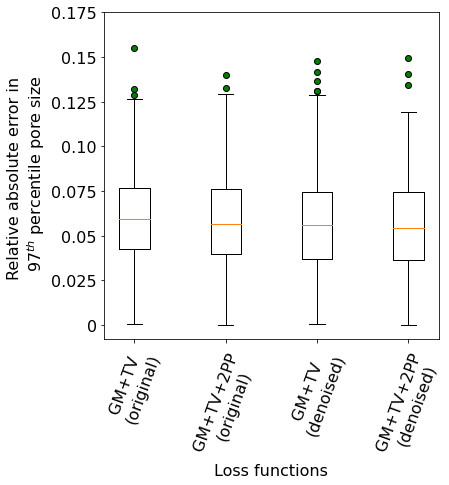}}\hfill
\caption{Errors in the statistical properties across all 185 slices}
  \label{fig:props_comp_all_slices}
\end{figure}
In this section, reconstruction is performed by considering all $185$ original microstructure slices and their corresponding denoised versions individually as the target images. Three different initializations of the reconstructed image (leading to three distinct statistically equivalent images) are chosen for each of target image. Therefore, overall there are 555 reconstructions cases and their results are shown in figure \ref{fig:props_comp_all_slices}. 
Figure \ref{fig:props_comp_all_slices} shows the comparison of the error in the statistical properties after reconstructions with the ``GM+TV" and ``GM+TV+2PP" loss cases, once with the original slices as the target microstructures and in the other instance, with the denoised slices as the target images. It is found that the comparison of almost all statistical properties are very similar to each other for all four reconstruction scenarios. The only exception is for the specific boundary length values where the overall error for the reconstruction cases with the denoised target slices are lower than that with the original target slices.
\section{FE analysis based material properties} \label{sec:FE analysis}
In this section, quality of the reconstructions are assessed in terms of a few averaged material properties of the porous ceramic material. The material properties of interest are the effective Young's modulus ($\bar{E}$), effective thermal conductivity ($\bar{\lambda}$) and effective hydraulic conductivity ($\bar{K}$). After an image is reconstructed based on a target image, the properties of the reconstructed microstructure are simulated using the commercial FE software ABAQUS \cite{hibbett1998abaqus}. $560 \times 902$ elements (each pixel assigned an element) are used to model each microstructure in 2D.
\begin{figure}[htpb]
  \centering
    \subfloat[Finite element model]{\label{fig:fe_model}\includegraphics[width=0.42\textwidth]{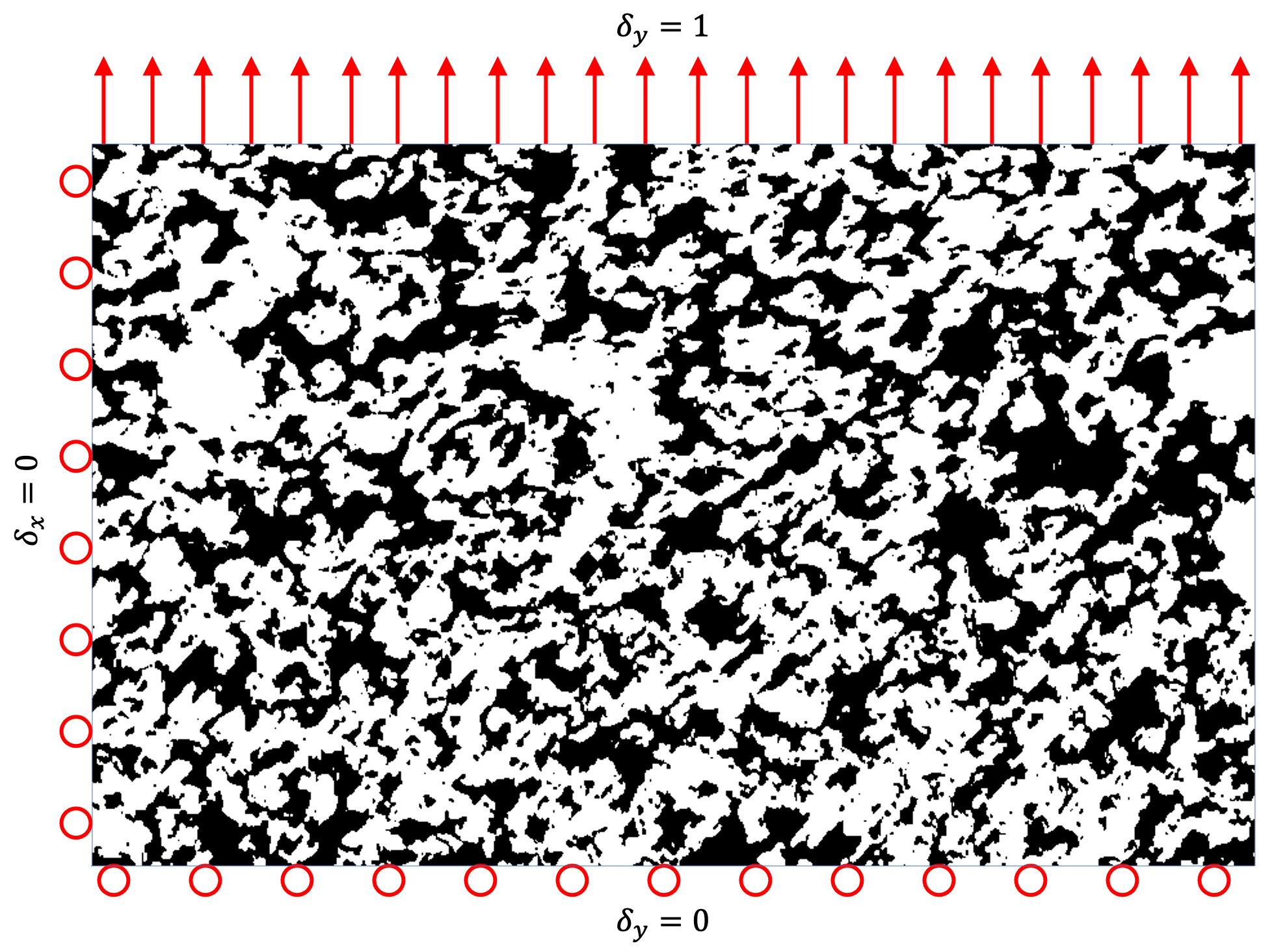}}\hfill
      \subfloat[Von Mises stress map]{\label{fig:EOI_Sec2}\includegraphics[width=0.495\textwidth]
{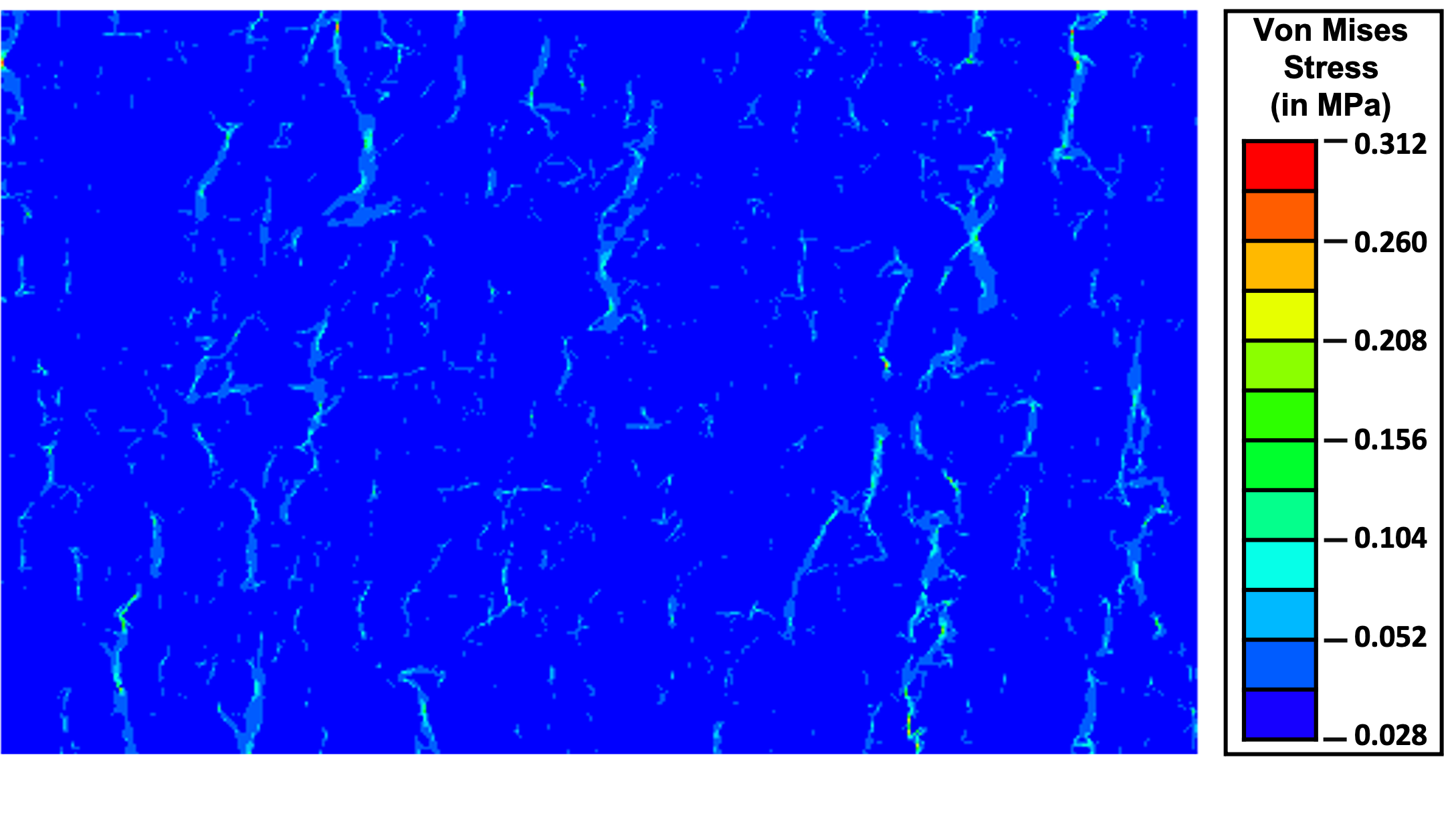}}\hfill
  \caption{Linear elastic analysis. (a) Finite element model: displacement is applied at the top face ($\delta_y=1$) with the bottom face held fixed ($\delta_y=0$), both along the vertical direction; the left face is also fixed along the horizontal direction ($\delta_x=0$); the in-plane Young's modulus for the solid phase is given by $E_{solid} = 100$ MPa, and that of the porous phase is given by $E_{pore} = 1$ MPa.  (b) Contour map of Von Mises stress obtained from the linear elastic analysis is shown here.}
  \label{fig:fe_model_stress_simulations}
\end{figure}

A static linear elastic analysis is performed to calculate the effective Young's modulus by subjecting the finite element model of the microstructure to a tensile load as shown in figure \ref{fig:fe_model_stress_simulations}. The macroscopic stress $\bar{\sigma}_{yy}$ in the vertical direction (direction of applied displacement) is calculated \cite{dalaq2016finite} by summing all the nodal reaction forces on the top face along the vertical direction and dividing it by the cross sectional area $A$ of the face:
\begin{equation}
\bar{\sigma}_{yy} = \frac{\sum_i^N F^R_i}{A}
\end{equation}
where $F^R_i$ is the reaction force on node $i$ along the vertical direction, $N$ is the number of nodes on the surface where displacement is applied. The effective Young's modulus is then obtained using:
\begin{equation}
\bar{E} = \frac{\bar{\sigma}_{yy}}{\bar{\epsilon}_{yy}}
\end{equation}
where macroscopic strain $\bar{\epsilon}_{yy}=\frac{\delta_y}{L}$ and $L$ is the edge length along the vertical direction.

\begin{figure}[h!]
  \centering
    \subfloat[Finite element model]{\label{fig:EOI_Sec1}\includegraphics[width=0.42\textwidth]
{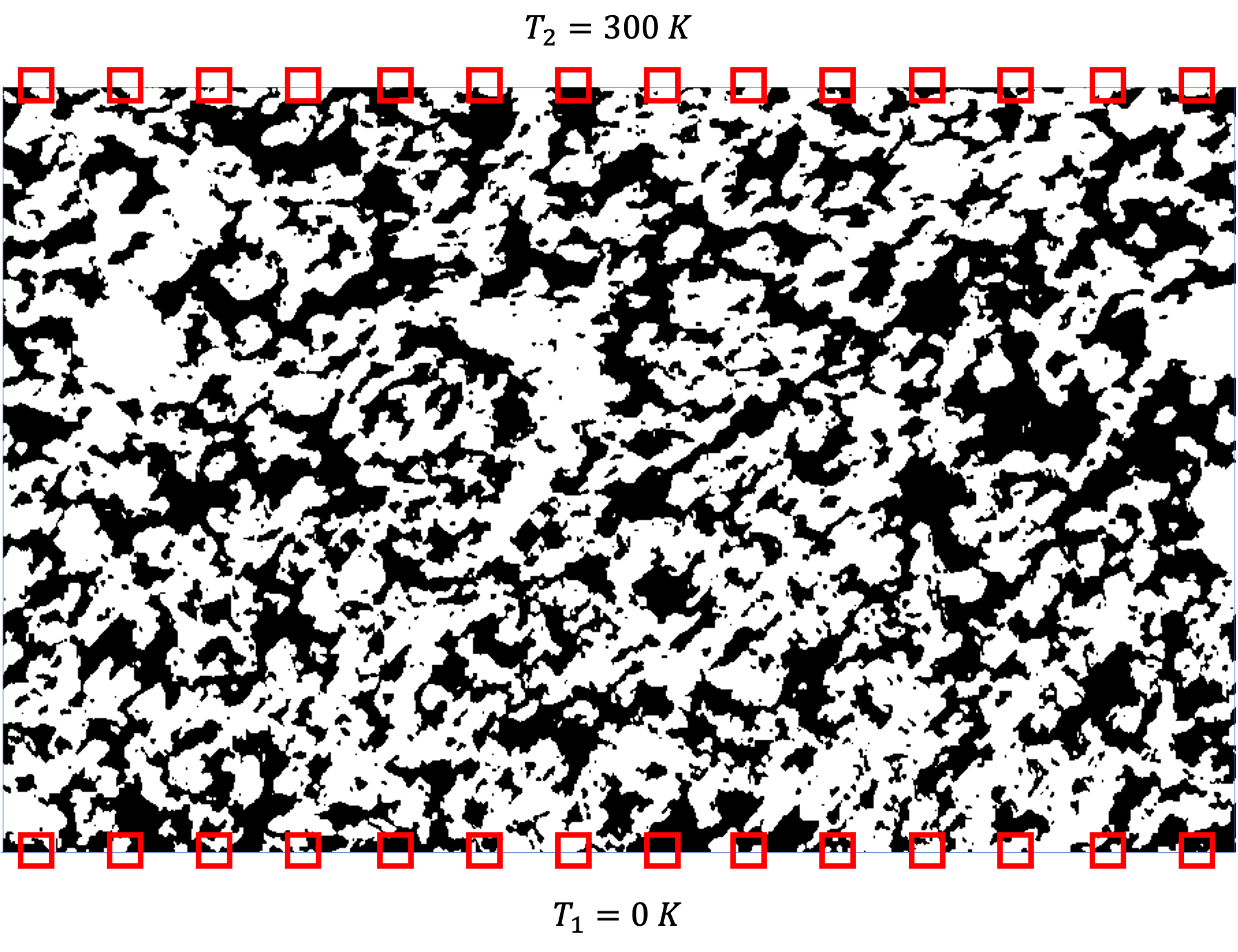}}\hfill
  \subfloat[Heat flux density map]{\label{fig:EOI_Sec2}\includegraphics[width=0.52\textwidth]
{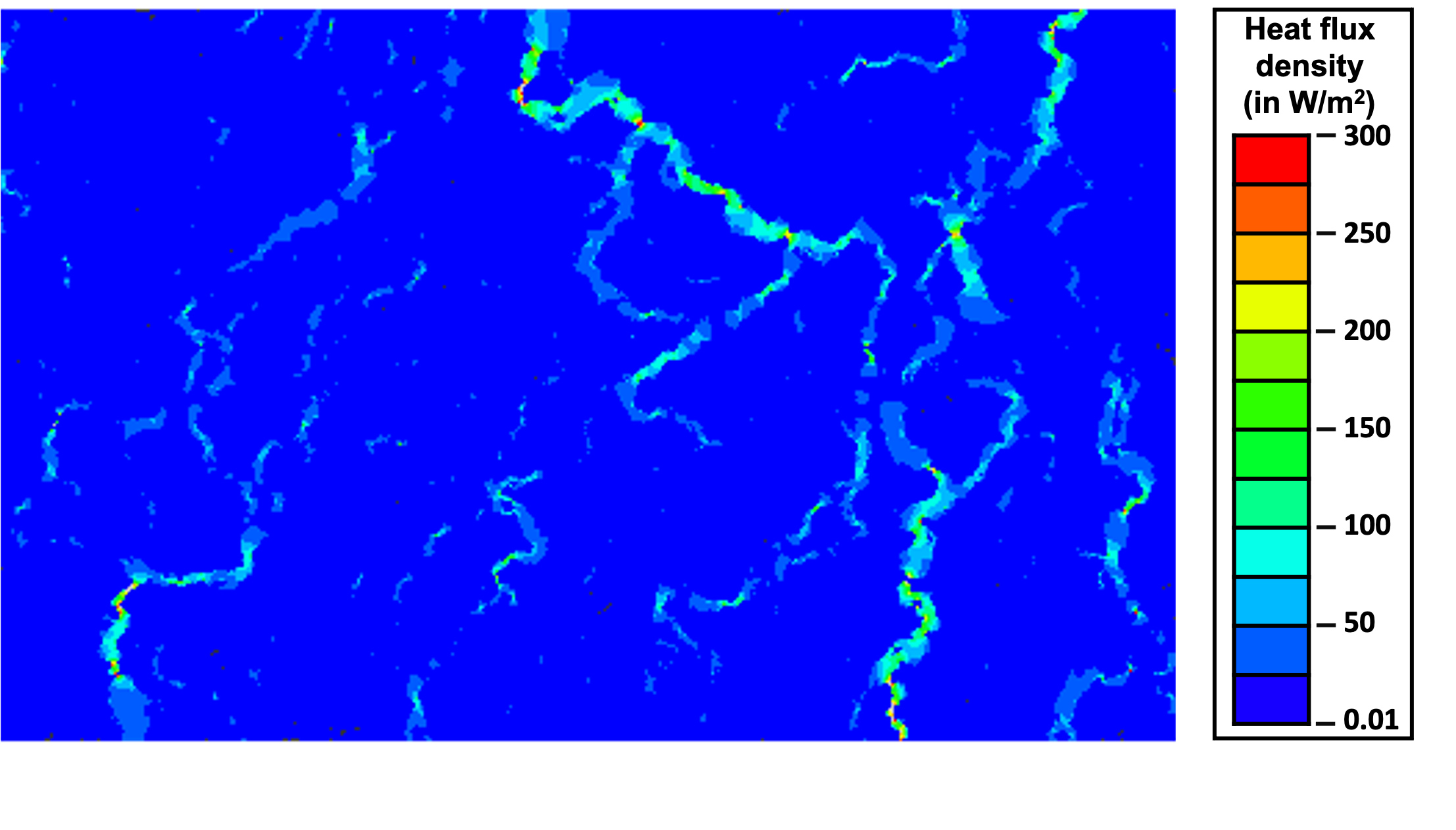}}\hfill
 \caption{Heat conduction analysis. (a) Finite element model: Temperature difference is applied between the top face ($T_2=300 \ K$) and the bottom face ($T_1=0 \ K$); the thermal conductivities for the solid phase and the porous phase are given by $\lambda_{solid} = 400$ W/m/K and $\lambda_{pore} = 1$ W/m/K respectively.  (b) Contour map of the heat flux density obtained from the heat conduction analysis is shown here.}
  \label{fig:fe_model_cond_simulations}
\end{figure}

The effective thermal conductivity is obtained by applying a constant temperature difference between the top and bottom faces of the finite element model of the microstructure \cite{ramani1995finite} as shown in figure \ref{fig:fe_model_cond_simulations}.  The macroscopic heat flux density $\bar{q}_y$ is calculated by summing the nodal heat fluxes on the top face. The effective thermal conductivity is then calculated using Fourier's law of heat conduction \cite{bergman2011fundamentals}:
\begin{equation}
\bar{\lambda} = \bar{q}_y \frac{L}{\Delta T}
\end{equation}
where $\Delta T= T_2 - T_1$ is the temperature difference between the top and bottom face, and $L$ is the length of the edge along the vertical direction. \\
\begin{figure}[h!]
  \centering
   \subfloat[Finite element model]{\label{fig:EOI_Sec2}\includegraphics[width=0.42\textwidth]
{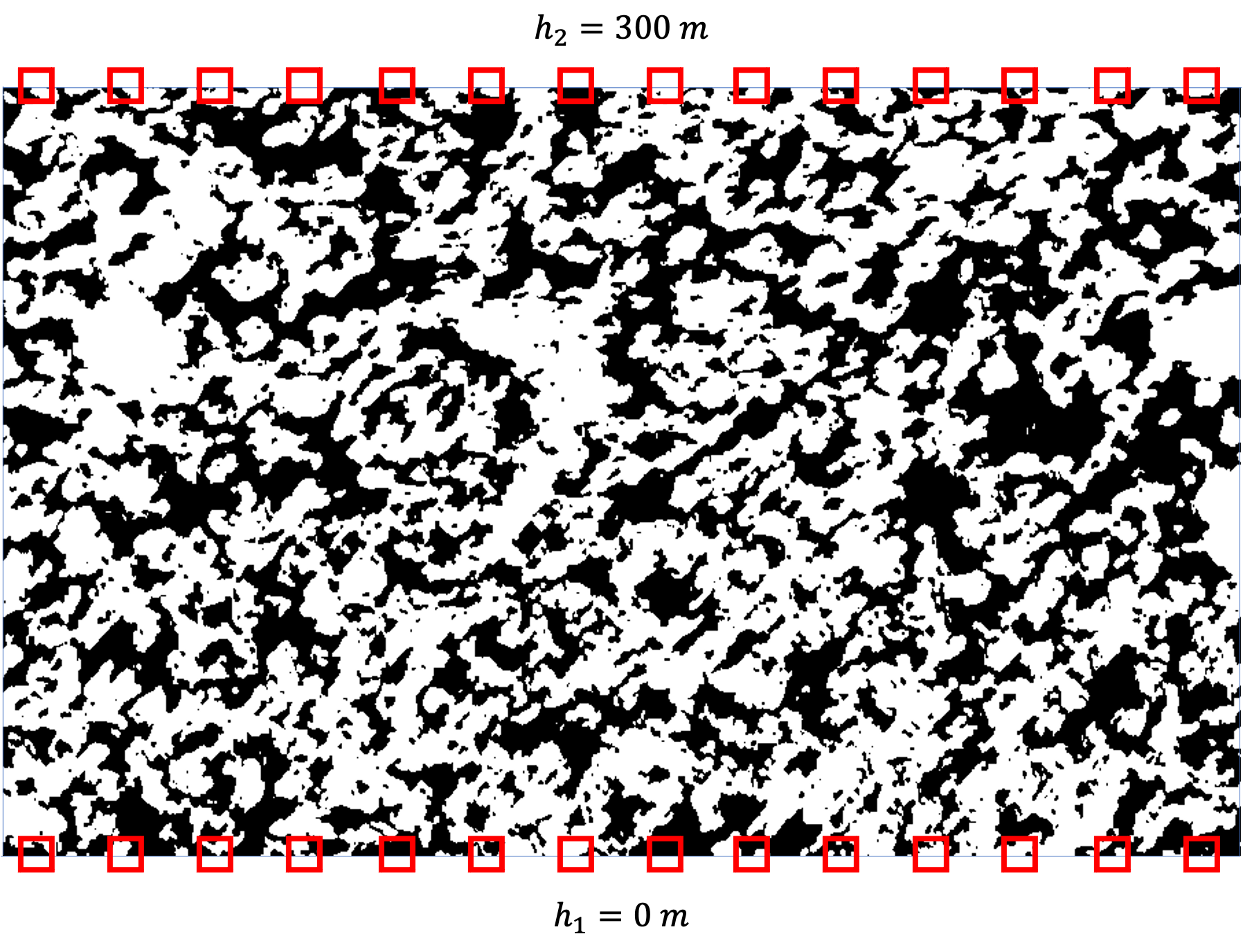}}\hfill
   \subfloat[Specific discharge map]{\label{fig:EOI_Sec2}\includegraphics[width=0.52\textwidth]
{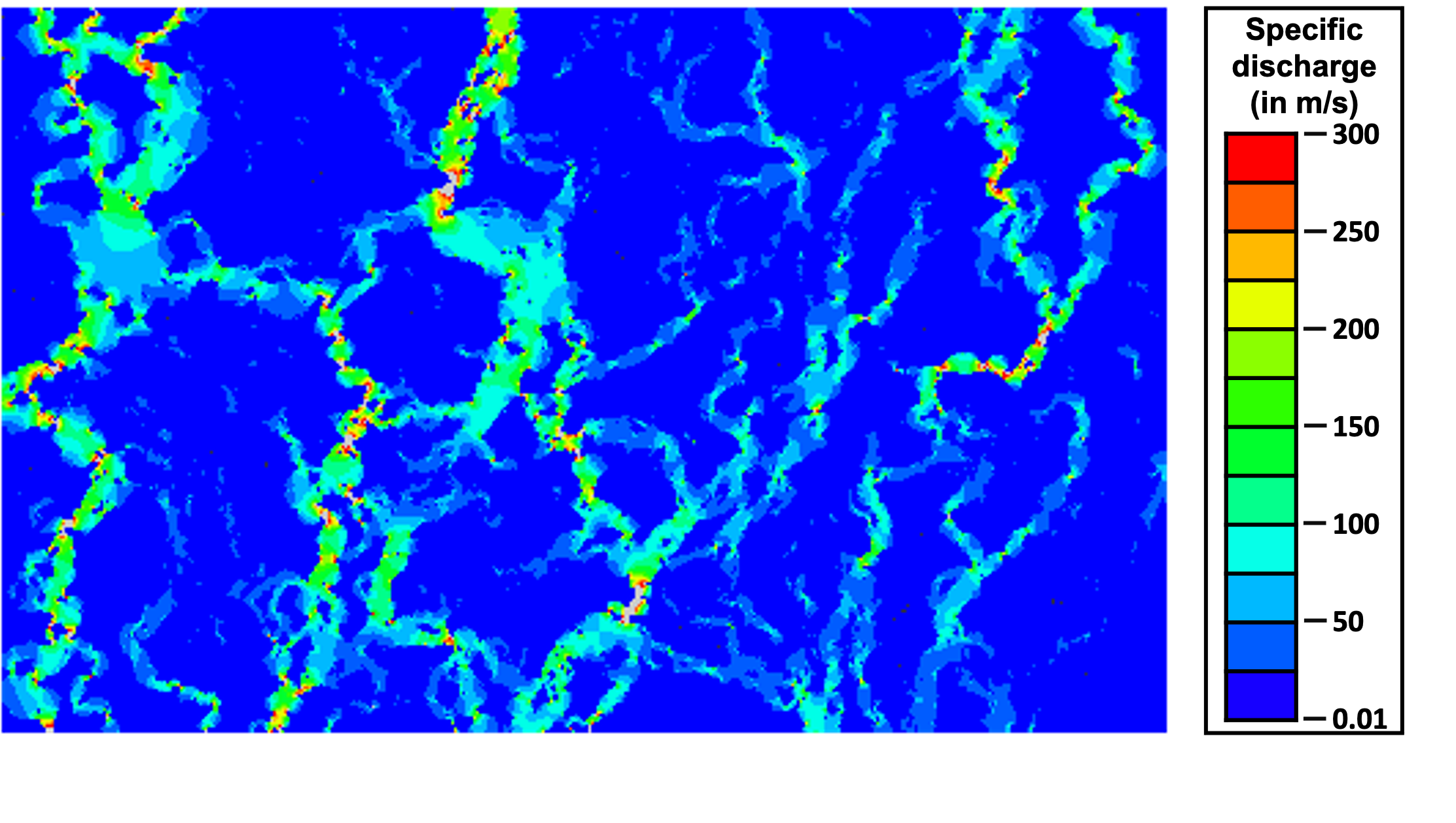}}\hfill
  \caption{Fluid flow analysis. (a) Finite element model: Hydraulic head is applied between the top face ($h_2=300 m$) and the bottom face ($h_1=0 m$); the hydraulic conductivities for the porous phase and the solid phase are given by $K_{solid} = 400$ m/s and $K_{pore} = 1$ m/s respectively.  (b) Contour map of the specific discharge obtained from the fluid flow analysis is shown here.}
  \label{fig:fe_model_perm_simulations}
\end{figure}
\indent The effective hydraulic conductivity for a laminar fluid flow through the porous medium of the material is simulated by specifying the hydraulic head between the top and bottom faces of the finite element model of the microstructure and then calculated from Darcy’s equation as shown in figure \ref{fig:fe_model_perm_simulations}. The macroscopic specific discharge $\bar{q}_y$ is calculated by summing up the nodal specific discharge values on the top face. The effective hydraulic conductivity is then calculated using Darcy's law of fluid flow \cite{whitaker1986flow}:
\begin{equation}
\bar{K} = \bar{q}_y \frac{L}{\Delta h}
\end{equation}
where $\Delta h= h_2 - h_1$ is the hydraulic head between the top and bottom face, and $L$ is the length of the edge along the vertical direction.
\begin{figure}[htpb]
  \centering
  \subfloat[Effective modulus]{\label{fig:EOI_Sur}\includegraphics[width=0.33\textwidth]{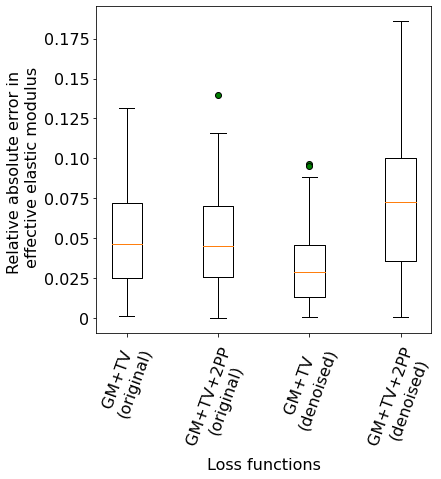}}\hfill
  \subfloat[Effective thermal \\ $ {} \ \ \ \ \ \ \ {}$ conductivity]{\label{fig:EOI_Sec1}\includegraphics[width=0.32\textwidth]
{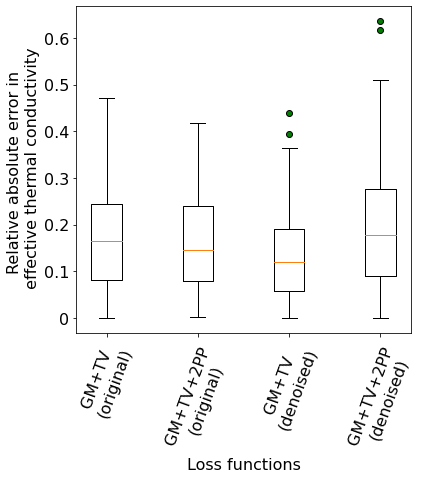}}\hfill
 \subfloat[Effective hydraulic \\ $ {} \ \ \ \ \ \ \ {}$ conductivity]{\label{fig:EOI_Sec2}\includegraphics[width=0.32\textwidth]
{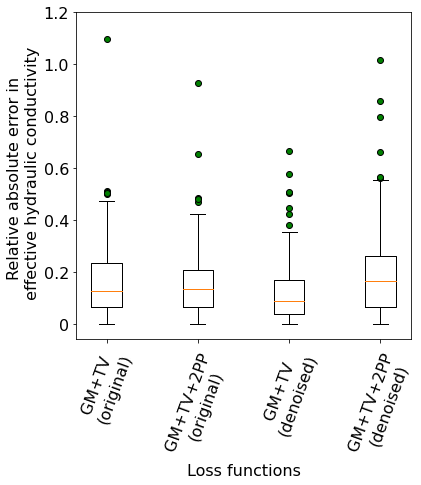}}\hfill
\caption{Errors in the material properties across all 185 slices}
  \label{fig:E_K_P_comp_all_slices}
\end{figure}

Figure \ref{fig:E_K_P_comp_all_slices} shows the boxplot of the relative absolute errors in the effective modulus, conductivity and hydraulic conductivity of the reconstructed microstructures with respect to the corresponding target microstructures. 
It is seen in figure \ref{fig:E_K_P_comp_all_slices}(a) that the effective Young's modulus error values for all the cases are in an acceptable range. The reconstruction quality is similar between the two loss function cases when the original images are used as target slices. However, when the denoised images are used, the ``GM+TV" loss case performs better than the ``GM+TV+2PP" case. The error values corresponding to the effective thermal conductivity in figure  \ref{fig:E_K_P_comp_all_slices}(b) and the effective hydraulic conductivity in figure \ref{fig:E_K_P_comp_all_slices}(c) are significantly higher than that of the elastic modulus. This suggests that the reconstruction algorithm is not able to capture these properties efficiently.  {\color{black} This can be attributed to the fact that the effective modulus is a function of the overall pattern of the microstructure, i.e., the relative arrangement of the black solid and the white porous phase, which is well matched in the original and the reconstructed microstructure. However, the effective thermal conductivity and hydraulic conductivity depend significantly on localizations in the microstructures that enable heat flow and fluid flow respectively. The goal of the reconstruction algorithm is to achieve statistical equivalence by matching the pattern and statistics of the original and reconstructed microstructure across the entire image domain. Thus,  even with a statistically equivalent reconstructed image, differences in small localized regions can lead to significant differences in the heat and fluid flow channels, resulting in very different effective thermal and hydraulic conductivity. Overall, if the error values are compared among all the four reconstruction cases for each material property, it is seen that the ``GM+TV" loss case with the denoised slices as the target images yields the most accurate reconstructions.}
{\color{black}
\begin{figure}[b!]
  \centering
    \subfloat[Original slice $\#$1 ($560\times902$ $\text{pixels}^2$)]{\label{fig:EOI_Sur}\includegraphics[width=0.4375\textwidth]{slice1_original.png}}\hfill
      \subfloat[Reconstructed image ($700\times1000$ $\text{pixels}^2$)]{\label{fig:EOI_Sec2}\includegraphics[width=0.5\textwidth]{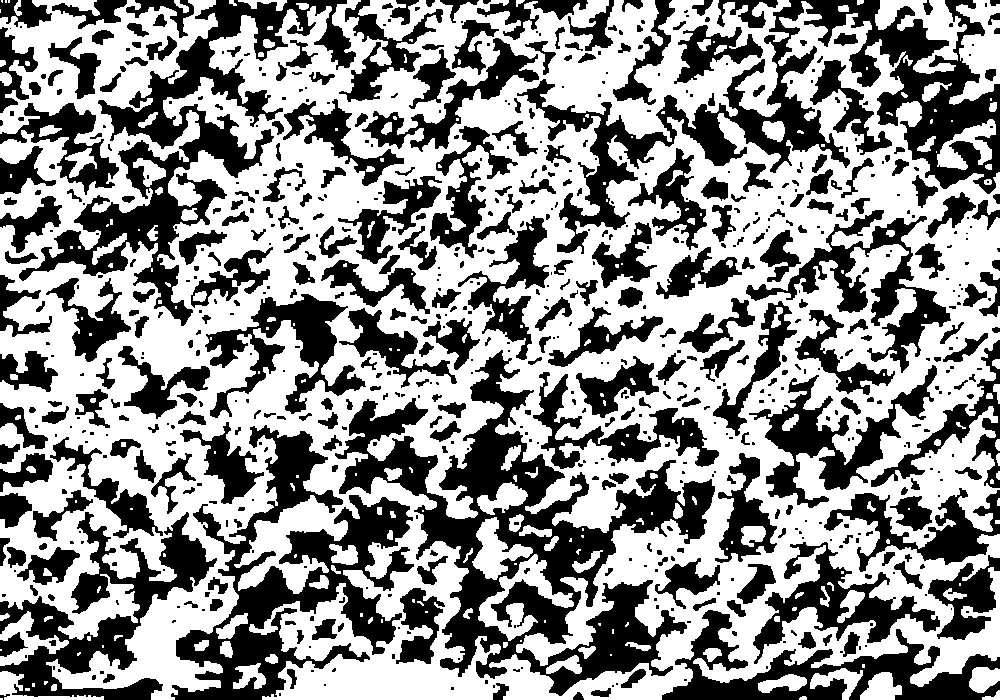}}\hfill
  \caption{Comparison of the original and a larger reconstructed microstructure}
  \label{fig:original_larger_comparison}
\end{figure}
\section{Discussion}
Results in the previous sections have shown the performance of the proposed reconstruction approach to be reasonably accurate and efficient. One further advantage of the proposed approach is that the reconstructed image is not constrained to be of the same size as the original image. Obtaining a high resolution image of a microstructure can be expensive, which limits the size and/or resolution of the obtained image. It is often of interest to analyze a microstructural image of dimensions larger than that of the original image. One of the useful features of this approach is the ability to reconstruct microstructure of spatial dimensions different from that of the original microstructural image. Here, as a demonstration, reconstruction of an image of size $700$ pixels $\times$ $1000$ pixels is considered using the ``GM+TV+2PP" loss. The larger reconstructed image is shown in figure \ref{fig:original_larger_comparison} along with the original slice $\#1$ microstructure for a visual comparison. To demonstrate the statistical equivalence of the larger reconstruction microstructure with its target image, the two-point probability function plot for the porous phase and the pore size complementary cdf plot for the two images are compared and shown in figure \ref{fig:properties_larger_comparison}.

\begin{figure}[t!]
  \centering
    \subfloat[Two-point probability function \\
     ${} \ \ \ \ \ {} \ \ \ \ \ {} $ for the porous phase]{\label{fig:EOI_Sur}\includegraphics[width=0.4\textwidth]{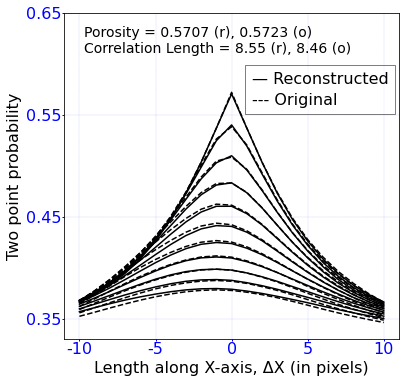}} \hfill
   \subfloat[Pore size complementary cdf]{\label{fig:EOI_Sec2}\includegraphics[width=0.4\textwidth]
{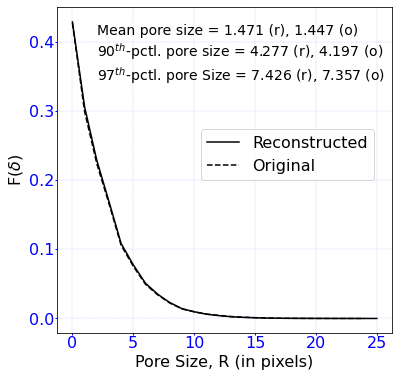}} \hfill
  \caption{Comparison of statistical properties between the original slice $\#1$ and the corresponding 700x1000 reconstructed microstructure using the ``GM+TV+2PP" loss function}
  \label{fig:properties_larger_comparison}
\end{figure}
\indent Another advantage of the proposed approach is that it is more efficient and accurate than other existing approaches based on statistical features of the microstructure. The approach is thus compared here with a modified version of the YT method \cite{yeong1998reconstructing, torquato1998reconstructing, graham2008stochastic} in terms of efficiency and accuracy. The YT method initiates by generating an image having the same porosity measure as the target microstructure.  It is an iterative method where in each iteration, a randomly chosen black solid phase pixel and a white porous phase pixel are interchanged in the image. If the pixel swapping leads to an improvement in the statistical measure (equivalent to a reduction in loss function measure) of the new image compared to the previous one, the new image is updated to be the current reconstructed image. Otherwise, the previous image remains the current reconstructed image at the end of that iteration. The pixel swapping is also allowed if a random number generator, generating random numbers uniformly between $0$ and $1$, returns a value greater than or equal to $p_I$. Here $p_I$ is defined as the pixel interchange probability. The iterative convergence of the reconstructed image to the target image is measured by the root of sum of squares error (RSSE) between the two-point correlation function of the reconstructed image and the target image. With increase in the number of iterations, RSSE has a decreasing trend and the reconstructed image converges closer to the target microstructure with respect to the two-point correlation function measure.  
\begin{figure}[t]
  \centering
    \subfloat[Original slice $\#1$]{\label{fig:YT_comparison_orig}\includegraphics[width=0.46\textwidth]{slice1_original.png}}\hfill \\
      \subfloat[Multi-loss based reconstruction]{\label{fig:YT_comparison_proposed}\includegraphics[width=0.46\textwidth]
{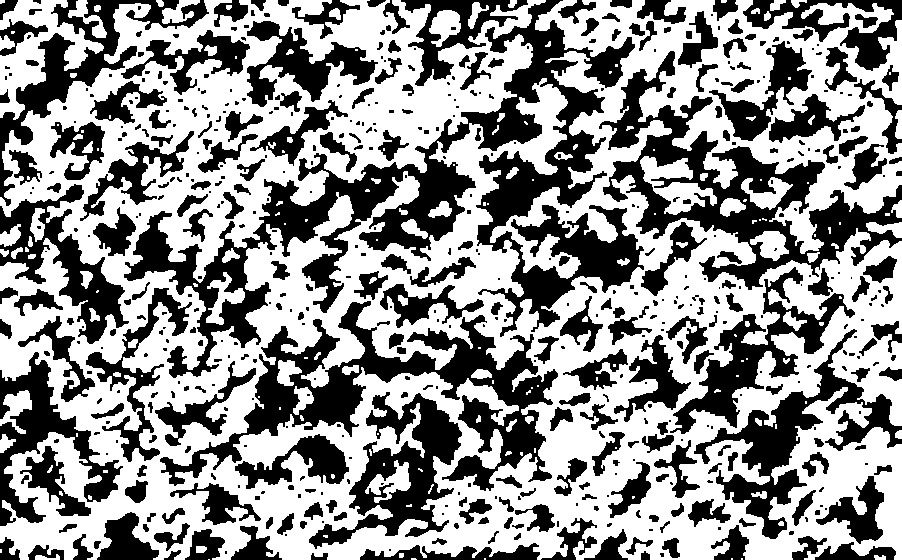}}\hfill
    \subfloat[YT method based reconstruction]{\label{fig:YT_comparison_YT}\includegraphics[width=0.46\textwidth]
{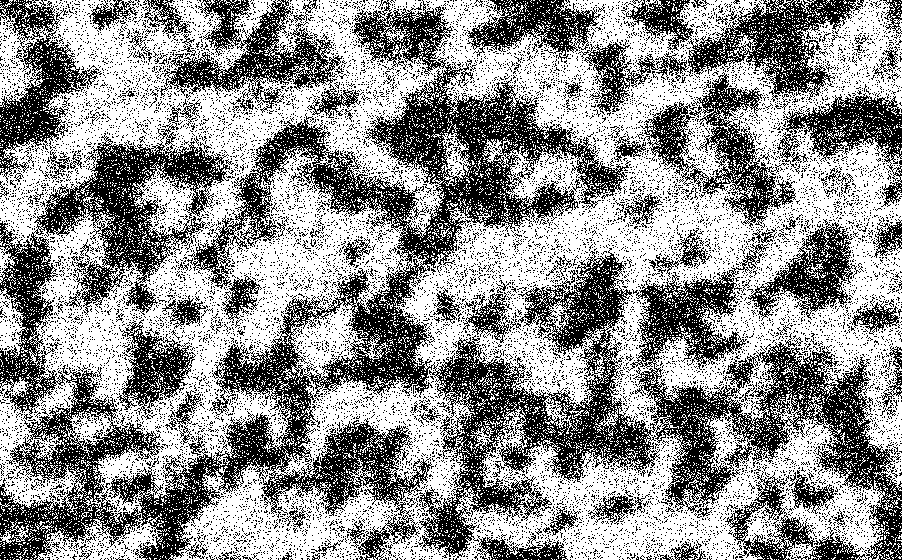}}\hfill
  \caption{Comparison of reconstructed images using the mutli-loss optimization based recontruction algorithm and a modified version of the YT method.}
  \label{fig:YT_comparison}
\end{figure}
Figure \ref{fig:YT_comparison} shows a visual comparison of the reconstructed images formed by using the proposed reconstruction approach and the YT method. Slice $\#1$, shown in figure \ref{fig:YT_comparison_orig}, is used as the target image in this case. It is clearly seen that the quality of the reconstructed image obtained by the proposed approach is better than that of the YT method. In this case, the YT method is run for $1$ million iterations and it takes around $90$ minutes to obtain the image in figure \ref{fig:YT_comparison_YT}. On the other hand, the proposed algorithm takes $17$,$400$ iterations and only around $18$ minutes to produce the image in figure \ref{fig:YT_comparison_proposed}. Thus, the proposed approach is found to be much more efficient and accurate than the YT method in the reconstruction of the porous ceramic material. {\color{black}It is noted that the algorithms are run on Google Colaboratory Pro \cite{bisong2019google}, a cloud-based Jupyter notebook environment.}
%
}
\section{Conclusions} \label{sec:conclusions}
In this paper, a modified version of an existing transfer learning reconstruction approach \cite{li2018transfer} is presented by introducing additional loss functions in the optimization framework. It is then used to reconstruct a binary porous ceramic material. A thorough statistical study was done to check the robustness of the algorithm over different initializations and over different microstructure slices (original and denoised) of the same material. Furthermore, the reconstruction quality is assessed not only based on the statistical properties of the microstructure but also the FEA simulated effective material properties. This algorithm has the advantage of being easily extended to 3D microstructure reconstruction \cite{sundararaghavan2005classification, bochenek2004reconstruction, bhandari20073d} if a suitable pretrained 3D convolutional network is available. It also has the flexibility of generating microstructures of size different from that of the target microstructure. A potential future work can be incorporating the simulated effective material property information within the reconstruction algorithm framework in an efficient manner, in order to generate microstructures with effective material properties closer to that of the target microstructure. 
\section*{Acknowledgements}
Research was sponsored by the Army Research Laboratory and was accomplished under Cooperative Agreement Number W911NF-12-2-0023 and W911NF-12-2-0022. The views and conclusions contained in this document are those of the authors and should not be interpreted as representing the official policies, either expressed or implied, of the Army Research Laboratory or the U.S. Government. The U.S. Government is authorized to reproduce and distribute reprints for Government purposes notwithstanding any copyright notation herein.
\bibliographystyle{ieeetr}
\bibliography{reference}

\begin{thebibliography}{10}
\expandafter\ifx\csname url\endcsname\relax
  \def\url#1{\texttt{#1}}\fi
\expandafter\ifx\csname urlprefix\endcsname\relax\def\urlprefix{URL }\fi
\expandafter\ifx\csname href\endcsname\relax
  \def\href#1#2{#2} \def\path#1{#1}\fi

\bibitem{hazlett1997statistical}
R.~Hazlett, Statistical characterization and stochastic modeling of pore
  networks in relation to fluid flow, Mathematical Geology 29~(6) (1997)
  801--822.

\bibitem{rintoul1997reconstruction}
M.~D. Rintoul, S.~Torquato, Reconstruction of the structure of dispersions,
  Journal of Colloid and Interface Science 186~(2) (1997) 467--476.

\bibitem{graham2008stochastic}
L.~Graham-Brady, X.~F. Xu, Stochastic morphological modeling of random
  multiphase materials, Journal of Applied Mechanics 75~(6) (2008).

\bibitem{collins2010three}
B.~Collins, K.~Matous, D.~Rypl, Three-dimensional reconstruction of
  statistically optimal unit cells of multimodal particulate composites,
  International Journal for Multiscale Computational Engineering 8~(5) (2010).

\bibitem{talukdar2002stochastic}
M.~Talukdar, O.~Torsaeter, M.~Ioannidis, J.~Howard, Stochastic reconstruction
  of chalk from 2d images, Transport in Porous Media 48~(1) (2002) 101--123.

\bibitem{gerke2012description}
K.~Gerke, M.~Karsanina, E.~Skvortsova, Description and reconstruction of the
  soil pore space using correlation functions, Eurasian Soil Science 45~(9)
  (2012) 861--872.

\bibitem{jiao2008modeling}
Y.~Jiao, F.~Stillinger, S.~Torquato, Modeling heterogeneous materials via
  two-point correlation functions. {II}. {A}lgorithmic details and
  applications, Physical Review E 77~(3) (2008) 031135.

\bibitem{tang2009pixel}
T.~Tang, Q.~Teng, X.~He, D.~Luo, A pixel selection rule based on the number of
  different-phase neighbours for the simulated annealing reconstruction of
  sandstone microstructure, Journal of Microscopy 234~(3) (2009) 262--268.

\bibitem{yeong1998reconstructing}
C.~Yeong, S.~Torquato, Reconstructing random media, Physical Review E 57~(1)
  (1998) 495.

\bibitem{torquato1998reconstructing}
S.~Torquato, C.~Yeong, Reconstructing random media. {II}. {T}hree-dimensional
  media from two-dimensional cuts, Physical Review E 58~(1) (1998) 224--233.

\bibitem{breneman2013stalking}
C.~M. Breneman, L.~C. Brinson, L.~S. Schadler, B.~Natarajan, M.~Krein, K.~Wu,
  L.~Morkowchuk, Y.~Li, H.~Deng, H.~Xu, Stalking the materials genome: A
  data-driven approach to the virtual design of nanostructured p olymers,
  Advanced Functional Materials 23~(46) (2013) 5746--5752.

\bibitem{xu2014descriptor1}
H.~Xu, Y.~Li, C.~Brinson, W.~Chen, A descriptor-based design methodology for
  developing heterogeneous microstructural materials system, Journal of
  Mechanical Design 136~(5) (2014).

\bibitem{xu2014descriptor2}
H.~Xu, D.~A. Dikin, C.~Burkhart, W.~Chen, Descriptor-based methodology for
  statistical characterization and 3{D} reconstruction of microstructural
  materials, Computational Materials Science 85 (2014) 206--216.

\bibitem{quiblier1984new}
J.~A. Quiblier, A new three-dimensional modeling technique for studying porous
  media, Journal of Colloid and Interface Science 98~(1) (1984) 84--102.

\bibitem{levitz1998off}
P.~Levitz, Off-lattice reconstruction of porous media: critical evaluation,
  geometrical confinement and molecular transport, Advances in Colloid and
  Interface Science 76 (1998) 71--106.

\bibitem{cahn1965phase}
J.~W. Cahn, Phase separation by spinodal decomposition in isotropic systems,
  The Journal of Chemical Physics 42~(1) (1965) 93--99.

\bibitem{yu2017characterization}
S.~Yu, Y.~Zhang, C.~Wang, W.-k. Lee, B.~Dong, T.~W. Odom, C.~Sun, W.~Chen,
  Characterization and design of functional quasi-random nanostructured
  materials using spectral density function, Journal of Mechanical Design
  139~(7) (2017).

\bibitem{schmidhuber2015deep}
J.~Schmidhuber, Deep learning in neural networks: An overview, Neural Networks
  61 (2015) 85--117.

\bibitem{lecun2015deep}
Y.~LeCun, Y.~Bengio, G.~Hinton, Deep learning, Nature 521~(7553) (2015)
  436--444.

\bibitem{cristianini2000introduction}
N.~Cristianini, J.~Shawe-Taylor, et~al., An introduction to support vector
  machines and other kernel-based learning methods, Cambridge University Press,
  2000.

\bibitem{williams1998prediction}
C.~K. Williams, Prediction with gaussian processes: From linear regression to
  linear prediction and beyond, in: Learning in Graphical Models, Springer,
  1998, pp. 599--621.

\bibitem{bhaduri2018efficient}
A.~Bhaduri, L.~Graham-Brady, An efficient adaptive sparse grid collocation
  method through derivative estimation, Probabilistic Engineering Mechanics 51
  (2018) 11--22.

\bibitem{bhaduri2018stochastic}
A.~Bhaduri, Y.~He, M.~D. Shields, L.~Graham-Brady, R.~M. Kirby, Stochastic
  collocation approach with adaptive mesh refinement for parametric uncertainty
  analysis, Journal of Computational Physics 371 (2018) 732--750.

\bibitem{bhaduri2020free}
A.~Bhaduri, J.~Gardner, C.~F. Abrams, L.~Graham-Brady, Free energy calculation
  using space filled design and weighted reconstruction: a modified single
  sweep approach, Molecular Simulation 46~(3) (2020) 193--206.

\bibitem{bhaduri2020usefulness}
A.~Bhaduri, D.~Brandyberry, M.~D. Shields, P.~Geubelle, L.~Graham-Brady, On the
  usefulness of gradient information in surrogate modeling: Application to
  uncertainty propagation in composite material models, Probabilistic
  Engineering Mechanics 60 (2020) 103024.

\bibitem{bhaduri2020probabilistic}
A.~Bhaduri, C.~S. Meyer, J.~W. Gillespie~Jr, B.~Z. Haque, M.~D. Shields,
  L.~Graham-Brady, Probabilistic modeling of discrete structural response with
  application to composite plate penetration models, arXiv preprint
  arXiv:2011.11780 (2020).

\bibitem{decost2017exploring}
B.~L. DeCost, T.~Francis, E.~A. Holm, Exploring the microstructure manifold:
  image texture representations applied to ultrahigh carbon steel
  microstructures, Acta Materialia 133 (2017) 30--40.

\bibitem{lubbers2017inferring}
N.~Lubbers, T.~Lookman, K.~Barros, Inferring low-dimensional microstructure
  representations using convolutional neural networks, Physical Review E 96~(5)
  (2017) 052111.

\bibitem{cang2017microstructure}
R.~Cang, Y.~Xu, S.~Chen, Y.~Liu, Y.~Jiao, M.~Yi~Ren, Microstructure
  representation and reconstruction of heterogeneous materials via deep belief
  network for computational material design, Journal of Mechanical Design
  139~(7) (2017).

\bibitem{lee2009convolutional}
H.~Lee, R.~Grosse, R.~Ranganath, A.~Y. Ng, Convolutional deep belief networks
  for scalable unsupervised learning of hierarchical representations, in:
  Proceedings of the 26th Annual International Conference on Machine Learning,
  2009, pp. 609--616.

\bibitem{li2018deep}
X.~Li, Z.~Yang, L.~C. Brinson, A.~Choudhary, A.~Agrawal, W.~Chen, A deep
  adversarial learning methodology for designing microstructural material
  systems, in: International Design Engineering Technical Conferences and
  Computers and Information in Engineering Conference, Vol. 51760, American
  Society of Mechanical Engineers, 2018, p. V02BT03A008.

\bibitem{goodfellow2014generative}
I.~Goodfellow, J.~Pouget-Abadie, M.~Mirza, B.~Xu, D.~Warde-Farley, S.~Ozair,
  A.~Courville, Y.~Bengio, Generative adversarial nets, in: Advances in Neural
  Information Processing Systems, 2014, pp. 2672--2680.

\bibitem{li2018transfer}
X.~Li, Y.~Zhang, H.~Zhao, C.~Burkhart, L.~C. Brinson, W.~Chen, A transfer
  learning approach for microstructure reconstruction and structure-property
  predictions, Scientific Reports 8~(1) (2018) 1--13.

\bibitem{simonyan2014very}
K.~Simonyan, A.~Zisserman, Very deep convolutional networks for large-scale
  image recognition, arXiv preprint arXiv:1409.1556 (2014).

\bibitem{deng2009imagenet}
J.~Deng, W.~Dong, R.~Socher, L.-J. Li, K.~Li, L.~Fei-Fei, Imagenet: A
  large-scale hierarchical image database, in: 2009 IEEE Conference on Computer
  Vision and Pattern Recognition, IEEE, 2009, pp. 248--255.

\bibitem{nickerson2019permeability}
S.~Nickerson, Y.~Shu, D.~Zhong, C.~K{\"o}nke, A.~Tandia, Permeability of porous
  ceramics by {X}-ray {CT} image analysis, Acta Materialia 172 (2019) 121--130.

\bibitem{kovci20193d}
P.~Ko{\v{c}}{\'\i}, M.~Isoz, M.~Plach{\'a}, A.~Arvajov{\'a},
  M.~V{\'a}clav{\'\i}k, M.~Svoboda, E.~Price, V.~Nov{\'a}k, D.~Thompsett, 3{D}
  reconstruction and pore-scale modeling of coated catalytic filters for
  automotive exhaust gas aftertreatment, Catalysis Today 320 (2019) 165--174.

\bibitem{torquato2002random}
S.~Torquato, H.~Haslach~Jr, Random heterogeneous materials: microstructure and
  macroscopic properties, Applied Mechanics Reviews 55~(4) (2002) B62--B63.

\bibitem{vanmarcke1983random}
E.~Vanmarcke, Random fields, 1983.

\bibitem{prager1963interphase}
S.~Prager, Interphase transfer in stationary two-phase media, Chemical
  Engineering Science 18~(4) (1963) 227--231.

\bibitem{aghdam2017guide}
H.~H. Aghdam, E.~J. Heravi, Guide to convolutional neural networks, New York,
  NY: Springer 10 (2017) 978--973.

\bibitem{venkatesan2017convolutional}
R.~Venkatesan, B.~Li, Convolutional neural networks in visual computing: a
  concise guide, CRC Press, 2017.

\bibitem{krizhevsky2017imagenet}
A.~Krizhevsky, I.~Sutskever, G.~E. Hinton, Imagenet classification with deep
  convolutional neural networks, Communications of the ACM 60~(6) (2017)
  84--90.

\bibitem{romanuke2017appropriate}
V.~Romanuke, Appropriate number and allocation of relus in convolutional neural
  networks, Scientific News of the National Technical University of Ukraine,
  Kyiv Polytechnic Institute~(1) (2017) 69--78.

\bibitem{ciregan2012multi}
D.~Ciregan, U.~Meier, J.~Schmidhuber, Multi-column deep neural networks for
  image classification, in: 2012 IEEE Conference on Computer Vision and Pattern
  Recognition, IEEE, 2012, pp. 3642--3649.

\bibitem{yamaguchi1990neural}
K.~Yamaguchi, K.~Sakamoto, T.~Akabane, Y.~Fujimoto, A neural network for
  speaker-independent isolated word recognition, in: First International
  Conference on Spoken Language Processing, 1990.

\bibitem{gatys2015texture}
L.~Gatys, A.~S. Ecker, M.~Bethge, Texture synthesis using convolutional neural
  networks, in: Advances in Neural Information Processing Systems, 2015, pp.
  262--270.

\bibitem{rudin1992nonlinear}
L.~I. Rudin, S.~Osher, E.~Fatemi, Nonlinear total variation based noise removal
  algorithms, Physica D: Nonlinear Phenomena 60~(1-4) (1992) 259--268.

\bibitem{hibbett1998abaqus}
Hibbett, Karlsson, Sorensen, ABAQUS/{S}tandard: User's Manual, Vol.~1, Hibbitt,
  Karlsson \& Sorensen, 1998.

\bibitem{dalaq2016finite}
A.~S. Dalaq, D.~W. Abueidda, R.~K.~A. Al-Rub, I.~M. Jasiuk, Finite element
  prediction of effective elastic properties of interpenetrating phase
  composites with architectured 3{D} sheet reinforcements, International
  Journal of Solids and Structures 83 (2016) 169--182.

\bibitem{ramani1995finite}
K.~Ramani, A.~Vaidyanathan, Finite element analysis of effective thermal
  conductivity of filled polymeric composites, Journal of Composite Materials
  29~(13) (1995) 1725--1740.

\bibitem{bergman2011fundamentals}
T.~L. Bergman, F.~P. Incropera, D.~P. DeWitt, A.~S. Lavine, Fundamentals of
  heat and mass transfer, John Wiley \& Sons, 2011.

\bibitem{whitaker1986flow}
S.~Whitaker, Flow in porous media {I}: {A} theoretical derivation of {D}arcy's
  law, Transport in Porous Media 1~(1) (1986) 3--25.

\bibitem{bisong2019google}
E.~Bisong, Google {C}olaboratory, in: Building Machine Learning and Deep
  Learning Models on Google Cloud Platform, Springer, 2019, pp. 59--64.

\bibitem{sundararaghavan2005classification}
V.~Sundararaghavan, N.~Zabaras, Classification and reconstruction of
  three-dimensional microstructures using support vector machines,
  Computational Materials Science 32~(2) (2005) 223--239.

\bibitem{bochenek2004reconstruction}
B.~Bochenek, R.~Pyrz, Reconstruction of random microstructures----a stochastic
  optimization problem, Computational Materials Science 31~(1-2) (2004)
  93--112.

\bibitem{bhandari20073d}
Y.~Bhandari, S.~Sarkar, M.~Groeber, M.~Uchic, D.~Dimiduk, S.~Ghosh, 3d
  polycrystalline microstructure reconstruction from fib generated serial
  sections for fe analysis, Computational Materials Science 41~(2) (2007)
  222--235.

\end{thebibliography}

\end{document}